\documentclass{aa}  
\usepackage{graphicx}
\bibpunct{(}{)}{;}{a}{}{,} 
\usepackage{txfonts}
\usepackage{ulem}
\begin{document}


\title{A star formation study of the ATLAS$^{\rm 3D}$ early-type
  galaxies with the AKARI all-sky survey}

\author{T.\ Kokusho\inst{1} \and H.\ Kaneda\inst{1} \and M.\
  Bureau\inst{2} \and T.\ Suzuki\inst{1} \and K.\ Murata\inst{1} \and
  A.\ Kondo\inst{1} \and M.\ Yamagishi\inst{3}}

\institute{Graduate School of Science, Nagoya University, Chikusa-ku,
  Nagoya 464-8602, Japan\\
  \email{kokusho@u.phys.nagoya-u.ac.jp}
  \and
  Sub-department of Astrophysics, Department of Physics, University of
  Oxford, Denys Wilkinson Building, Keble Road, Oxford OX1 3RH, UK\\
  \and 
  Institute of Space and Astronautical Science, Japan Aerospace
  Exploration Agency, Chuo-ku, Sagamihara 252-5210, Japan}


\abstract
{The star formation properties of early-type galaxies (ETGs) are
  currently the subject of considerable interest, particularly whether
  they differ from those of gas-rich spirals.}
{We perform a systematic study of star formation in a large sample of
  local ETGs using polycyclic aromatic hydrocarbon (PAH) and dust
  emission, focusing on the galaxies' star formation rates (SFRs) and
  star formation efficiencies (SFEs).}
{Our sample is composed of the $260$ ETGs from the ATLAS$^{\rm 3D}$
  survey, from which we use the cold gas measurements (\ion{H}{i} and
  CO). The SFRs are estimated from stellar, PAH and dust fits to
  spectral energy distributions created from new AKARI measurements
  and literature data from WISE and 2MASS.}
{The mid-infrared luminosities of non-CO-detected galaxies are well
  correlated with their stellar luminosities, showing that they trace
  (circum)stellar dust emission. CO-detected galaxies show an excess
  above these correlations, uncorrelated with their stellar
  luminosities, indicating that they likely contain PAHs and dust of
  interstellar origin. PAH and dust luminosities of CO-detected
  galaxies show tight correlations with their molecular gas masses,
  and the derived current SFRs are typically
  $0.01$\,--\,$1$~$M_\sun$~yr$^{-1}$. These SFRs systematically
  decrease with stellar age at fixed stellar mass, while they
  correlate nearly linearly with stellar mass at fixed age.  The
  majority of local ETGs follow the same star-formation law as local
  star-forming galaxies, and their current SFEs do not depend on
  either stellar mass or age.}
{Our results clearly indicate that molecular gas is fueling current
  star formation in local ETGs, that appear to acquire this gas via
  mechanisms regulated primarily by stellar mass. The current SFEs of
  local ETGs are similar to those of local star-forming galaxies,
  indicating that their low SFRs are likely due to smaller cold gas
  fractions rather than a suppression of star formation.}


\titlerunning{Star formation study of the ATLAS$^{\rm 3D}$ ETGs
  with AKARI}
\authorrunning{Kokusho et al.}

\maketitle


\section{Introduction}
\label{sec:intro}

Stellar population studies of local early-type galaxies (ETGs;
ellipticals and lenticulars) show that they are dominated by old stars
and reside in the so-called red sequence of the optical
colour-magnitude diagram. This suggests that they ceased to form stars
at an early stage of their evolution \citep[e.g.,][]{cow96,tho10}, and
thus represent the end point of galaxy evolution.

The interstellar space of ETGs is generally filled with hot X-ray gas
\citep[e.g.,][]{for85}, that can heat and destroy any cold
interstellar medium (ISM), the fuel for star formation. Nevertheless,
atomic gas, molecular gas and dust have all been detected in a
significant fraction of ETGs \citep[e.g.,][]{kna85,war86,kna89,kna96},
and evidence for the presence of such cold gas and dust in ETGs has
grown for the last two decades through both radio and infrared (IR)
observations. In particular, \citet{you11} carried out a molecular gas
survey of the $260$ ETGs in the ATLAS$^{\rm 3D}$ sample \citep{cap11},
that aimed to reveal the formation and evolutionary processes of
nearby ETGs. With a high detection rate of $22\%$, they confirmed that
molecular gas is prevalent in ETGs, suggesting that some ETGs may
still be able to form new stars. Far-infrared (FIR) emission from cold
dust in ETGs was systematically investigated first with the Infrared
Astronomical Satellite (IRAS; \citealt{kna89}). More recent
observations with satellites such as the Infrared Space Observatory
(ISO), Spitzer, AKARI and Herschel also detected dust of both
circumstellar and interstellar origin in many ETGs
\citep[e.g.,][]{tem03,kan11,smi12}. Polycyclic aromatic hydrocarbons
(PAHs), that are destroyed by X-rays much more easily than dust, are
also detected in some ETGs \citep[e.g.,][]{kan05,kan08,ram13}. As PAHs
are considered a star formation tracer, their presence in ETGs again
suggests that star formation is ongoing. Indeed, PAHs are excited and
ionised by far-ultraviolet (FUV) light from young stars, 
producing strong features primarily at wavelengths of $6.2$, $7.7$
and $8.6$~$\mu$m, while they are destroyed by hard UV and X-ray
radiation from active galactic nuclei (AGN; e.g., \citealt{pee04}).

Utilizing PAH emission observed with Spitzer, \citet{sha10}
investigated the star formation properties of the $48$ ETGs in the
SAURON sample \citep{zee02}, and argued that ETGs are forming stars
with surface densities and efficiencies similar to those of spiral
galaxies. They also suggested that star formation in ETGs proceeds in
at least two different modes (widespread and circumnuclear), that may
reflect different evolutionary paths. With PAH and mid-infrared
emission (MIR), \citet{cro11} calculated the current star formation
rates (SFRs) of the SAURON ETGs, and suggested that they form stars
less efficiently than spiral galaxies. Using both numerical
simulations and observational results, \citet{mar13} similarly argued
that ETGs have lower star formation efficiencies (SFEs; defined as the
SFR per unit cold gas mass $M_{\rm gas}$) than spiral galaxies, likely
due to the increased stability of discs embedded in spheroids
(morphological quenching). The same trend is reported in the
ATLAS$^{\rm 3D}$ sample using SFRs estimated from MIR and FUV light
\citep{dav14}. It has also been suggested that AGN and galaxy mergers
can respectively heat (and expel) and strip away the cold ISM of ETGs,
thus suppressing star formation  (i.e., lowering the SFE; see,
e.g., \citealt{sch06,nes11,ala15,gui15,lan16}). It is therefore
likely that while ETGs have a substantial cold ISM and are forming
stars, their SFE is reduced with respect to that of spiral galaxies,
due to a number of properties and mechanisms particular to them.

In a different vein, the sources of the cold ISM in ETGs are
themselves a matter of debate. As mentioned above, the stellar
populations of ETGs are uniformly old, indicating that the original
cold gas reservoirs that led to galaxy formation have already been
consumed \citep[e.g.,][]{bow92}. In addition, old stars do not
efficiently produce dust, and ETGs are filled with hot X-ray gas, that
can destroy dust through sputtering \citep[e.g.,][]{dra79}. Yet
\citet{gou95} clearly demonstrated that ellipticals contain more dust
than that expected from the balance between production via stellar
mass loss \citep[e.g.,][]{fab76} and destruction via X-ray
sputtering.  Galaxy mergers are therefore frequently invoked as a
possible mechanism to supply cold gas and dust to ETGs. For example,
\citet{dav11} studied the kinematics of stars, ionised gas and
molecular gas in ETGs and, due to the pervasiveness of kinematic
misalignments between stars and gas, suggested that the majority of
them may obtain cold gas from external sources. Internal sources are
however also proposed by several authors. For example, based on the
spatial distribution of dust and X-ray gas, it has been suggested that
intermittent buoyant outflows from low-luminosity AGN can replenish
dust in ellipticals \citep[e.g.,][]{tem07,kan11}. Continuous dust
production by old stars and dust growth in the ISM are other potential
channels for dust production in ETGs \citep{mar13_2,hir15}.

Depending on how ETGs obtain their gas and dust, star formation may
well proceed differently. A systematic study of the cold ISM
properties of a large sample of ETGs, coupled with constraints on
their star formation activity, is thus essential to properly
understand ETG evolution. Here, we systematically investigate the star
formation properties of the ATLAS$^{\rm 3D}$ ETGs using PAH and dust
emission measured from the AKARI all-sky survey, combined with
Wide-field Infrared Survey Explorer (WISE) and Two Micron All Sky
Survey (2MASS) archival data, and atomic and molecular gas data from
ATLAS$^{\rm 3D}$. We detail our measurements in Sect.~\ref{sec:data}
and present the main results in Sect.~\ref{sec:results}. The star
formation properties of ETGs are discussed in
Sect.~\ref{sec:discussion} and we summarise our results in
Sect.~\ref{sec:conclusions}.


\section{Sample and Data}
\label{sec:data}

\subsection{ATLAS$^{\rm 3D}$ survey}
\label{sec:atlas3d}

The goal of the ATLAS$^{\rm 3D}$ project was to conduct a
comprehensive study of a complete, volume-limited sample of $260$
morphologically-selected local ETGs ($M_{\rm K}<-21.5$ and $D<42$~Mpc;
\citealt{cap11}). A diverse range of observations and simulations was
performed, at its core optical integral-field spectroscopic
observations with the SAURON instrument \citep{bac01}, yielding
detailed information on the stellar kinematics
\citep{kra11,ems11,kra13,kra13b} and populations
\citep{cap12,sco13,mcd14,mcd15}. In addition to single-dish molecular
gas surveys \citep{com07,you11}, both \ion{H}{I} and CO
interferometric observations were obtained
\citep{you08,cro11,ser12,ala13}. The morphology and kinematics of
the molecular gas are characterised in \citet{dav13}, while its
physical conditions are discussed in \citet{cro12} and \citet{bay13}.

\subsection{AKARI all-sky survey}
\label{sec:akari}

The new IR fluxes presented here were obtained from the AKARI
satellite \citep{mur07} all-sky surveys in the MIR (central
wavelengths $9$ and $18$~$\mu$m; \citealt{ona07}) and FIR ($65$, $90$
and $140$~$\mu$m; \citealt{kaw07}). For the MIR $9$ and $18$~$\mu$m
bands, the total galaxy flux densities were obtained through aperture
photometry as follows. First, we applied a spatial filtering of
$1.3$~arcmin$^{-1}$ to each galaxy image, to remove periodic
noise. Second, we performed aperture photometry on the all-sky diffuse
maps (\citealt{doi15}; Ishihara et al., in preparation), within a
circular aperture of radius
$R_{\rm aper}=\sqrt{(2R_{\rm e})^2+(1.5D_{\rm PSF})^2}$, where
$R_{\rm e}$ is the effective radius of each galaxy in the optical $B$
band \citep{cap11} and $D_{\rm PSF}$ is the full width at half maximum
of the point spread function (PSF) of the AKARI instrument at the
given wavelength \citep{ish10,tak15}. The average sky level was
  first measured in a circular annulus of inner radius
  $1.5R_{\rm aper}$ and outer radius $2.5R_{\rm aper}$, and was then
  subtracted from individual measurements. The uncertainty of the
total flux density of a given object in a given band was estimated by
appropriately propagating the uncertainty on individual measurements,
taken as the standard deviation of the flux densities in the same
sky annulus. For the FIR $65$, $90$ and $140$~$\mu$m bands, the
total galaxy flux densities were measured in an identical manner, but
without the initial spatial filtering. Details of the measurement
will be provided in the AKARI ETG catalogue (Kaneda et al., in
preparation), in which integrated flux density measurements of
${\approx}8000$ ETGs selected from the HyperLeda catalogue
\citep{mak14} will be detailed.

Overall, we measured AKARI flux densities for $258$, $260$, $245$,
$246$ and $254$ of the $260$ ATLAS$^{\rm 3D}$ ETGs in the $9$, $18$,
$65$, $90$ and $140$~$\mu$m bands, respectively, the missing
measurements being due to incomplete AKARI sky coverage. When
possible, the missing $65$ and $90$~$\mu$m flux densities were
replaced by IRAS $60$ and/or $100$~$\mu$m measurements, as listed in
the NASA/IPAC Extragalactic Database
(NED)\footnote{https://ned.ipac.caltech.edu/}. In the $65$, $90$ and
$140$~$\mu$m bands, respectively, a further $52$, $8$ and $22$
measurements are likely affected by source confusion or instrumental
artefacts, and are thus excluded from our analysis. The MIR and FIR
flux densities thus derived are listed in Table~\ref{tab:fluxes}.

It has already been shown that AKARI fluxes are in good agreement
with analogous measurements by Spitzer and Herschel
\citep{suz10,hat16}, but we nevertheless check the quality of our
measurement here by comparing our AKARI $90$ and $140$~$\mu$m flux
densities with Spitzer and Herschel measurements. Among our sample,
$89$ galaxies have published Spitzer $70$~$\mu$m measurements
\citep{amb14}, and $22$ galaxies have both Herschel $100$ and
$160$~$\mu$m measurements \citep{dal12,ser13,bae14}. In all cases
(AKARI $90$~$\mu$m vs.\ Spitzer $70$~$\mu$m, AKARI $90$~$\mu$m vs.\
Herschel $100$~$\mu$m and AKARI $140$~$\mu$m vs.\ Herschel
$160$~$\mu$m), we find that most measurements are consistent within
the uncertainties, and that they are proportional with a slope of
approximately unity and a small intrinsic scatter of 0.2 dex, thus
confirming the reliability of our measurements. These parameters are
estimated using a linear fit with free slope and nul intercept in
log-log space, adjusting the intrinsic scatter until the reduced
$\chi^2$ is equal to $1$ \citep[see][]{wil10}.

\begin{figure}
  \centering
  \includegraphics[width=\hsize]{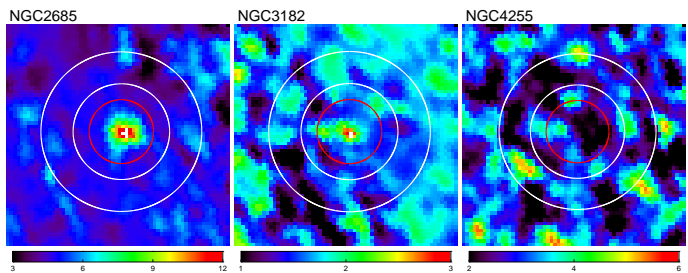}
  \caption{Examples of galaxies with a strong detection ($S/N>3$;
      left), a marginal detection ($S/N\approx3$; middle) and a
      non-detection ($S/N<3$; right) in the $140$~$\mu$m band. All
      panels are $16^{\prime}\times16^{\prime}$. Red circles and white
      annuli show the photometry and sky apertures,
      respectively. Color bars are in units of MJy~sr$^{-1}$.}
  \label{fig:image}
\end{figure}

The number of galaxies robustly detected (defined here as having a
signal-to-noise ratio $S/N>3$) is summarised in Table~\ref{tab:stats}
for each band. Detection rates are $41$, $13$, $16$, $43$ and $28\%$
for the $9$, $18$, $65$, $90$ and $140$~$\mu$m bands, respectively.
In the FIR, 117 galaxies ($45\%$) are detected in at least one band
(thus one, two or three FIR bands), while 71 galaxies ($27\%$) are
detected in at least two bands (thus two or three FIR bands).  The
lower detection rates at $18$ and $65$~$\mu$m are due to respectively
a relatively high background level \citep{kon16} and a narrow
bandwidth \citep{kaw07}. Systematic flux density uncertainties are
estimated to be $10\%$ in the MIR ($9$ and $18$~$\mu$m; Ishihara et
al., in preparation), and $20$, $20$ and $50\%$ in the FIR ($65$, $90$
and $140$~$\mu$m, respectively; \citealt{tak15}). For
illustrative purposes, Figure~\ref{fig:image} shows an example of a
strong detection ($S/N>3$), a marginal detection ($S/N\approx3$) and
a non-detection ($S/N<3$), all in the AKARI $140$~$\mu$m band.

\begin{table}
  \caption{AKARI robust detection ($S/N>3$) statistics.}
  \label{tab:stats}
  \centering
  \begin{tabular}{rcccc}
    \hline\hline
    Band & Detected & Not detected & Not observed & Total \\
    \hline
    9~$\mu$m & 106 & 152 & \phantom{1}2 & 260\\
    18~$\mu$m & \phantom{1}33 & 227 & \phantom{1}0 & 260\\
    65~$\mu$m & \phantom{1}42 & 217 & \phantom{1}1 & 260\\
    90~$\mu$m & 111 & 148 & \phantom{1}1 & 260\\
    140~$\mu$m & \phantom{1}70 & 184 & \phantom{1}6 & 260\\
    \hline
  \end{tabular}
  \tablefoot{The $14$ and $13$ galaxies not observed with AKARI in
    respectively the $65$ and $90$~$\mu$m bands are supplemented by 
    IRAS $60$ and/or $100$~$\mu$m measurements.}
\end{table}

\subsection{WISE survey}
\label{sec:wise}

Near-infrared (NIR) and MIR flux densities for our sample galaxies are
also available from the AllWISE catalog \citep{cut13}, composed of
data obtained during the cryogenic and post-cryogenic phases of the
WISE all-sky survey. We therefore use the four WISE photometric bands
with central wavelengths of $3.4$, $4.6$, $12$ and $22$~$\mu$m. Total
magnitudes were measured using aperture photometry (and are referred
to as WXGMAG, with X equal to 1, 2, 3 or 4 for each of the four WISE
bands), with the aperture calculated from 2MASS NIR surface photometry
\citep{skr06} and the WISE PSF. When a WXGMAG magnitude is not
available, we adopt instead the profile fit magnitude WXMPRO. However,
as some galaxies are much more spatially extended than the WISE PSF,
the profile fit photometry can underestimate the true total fluxes.
To avoid this, the WXMPRO magnitudes are calibrated using the method
of \citet{gri15}, that takes into account the extent of the galaxies
as measured from other WISE photometric parameters. Despite this, we
use the uncorrected profile fit magnitudes when there is strong
contamination from background sources, as the correction process is
then likely to be inaccurate. The 2MASS-determined WXGMAG apertures of
galaxies that are well resolved spatially can also be slightly too
small, resulting in total flux densities underestimated by as much as
$30\%$\footnote{http://wise2.ipac.caltech.edu/docs/release/allsky/expsup/sec6{\_}3e.html}.
We therefore add a systematic uncertainty of $30\%$ to the WISE
systematic flux calibration uncertainties.


\section{Results}
\label{sec:results}

\subsection{Relation between IR and $K_{\rm s}$-band luminosities}
\label{sec:ir-k}

\begin{figure*}
  \centering
  \includegraphics[width=11cm]{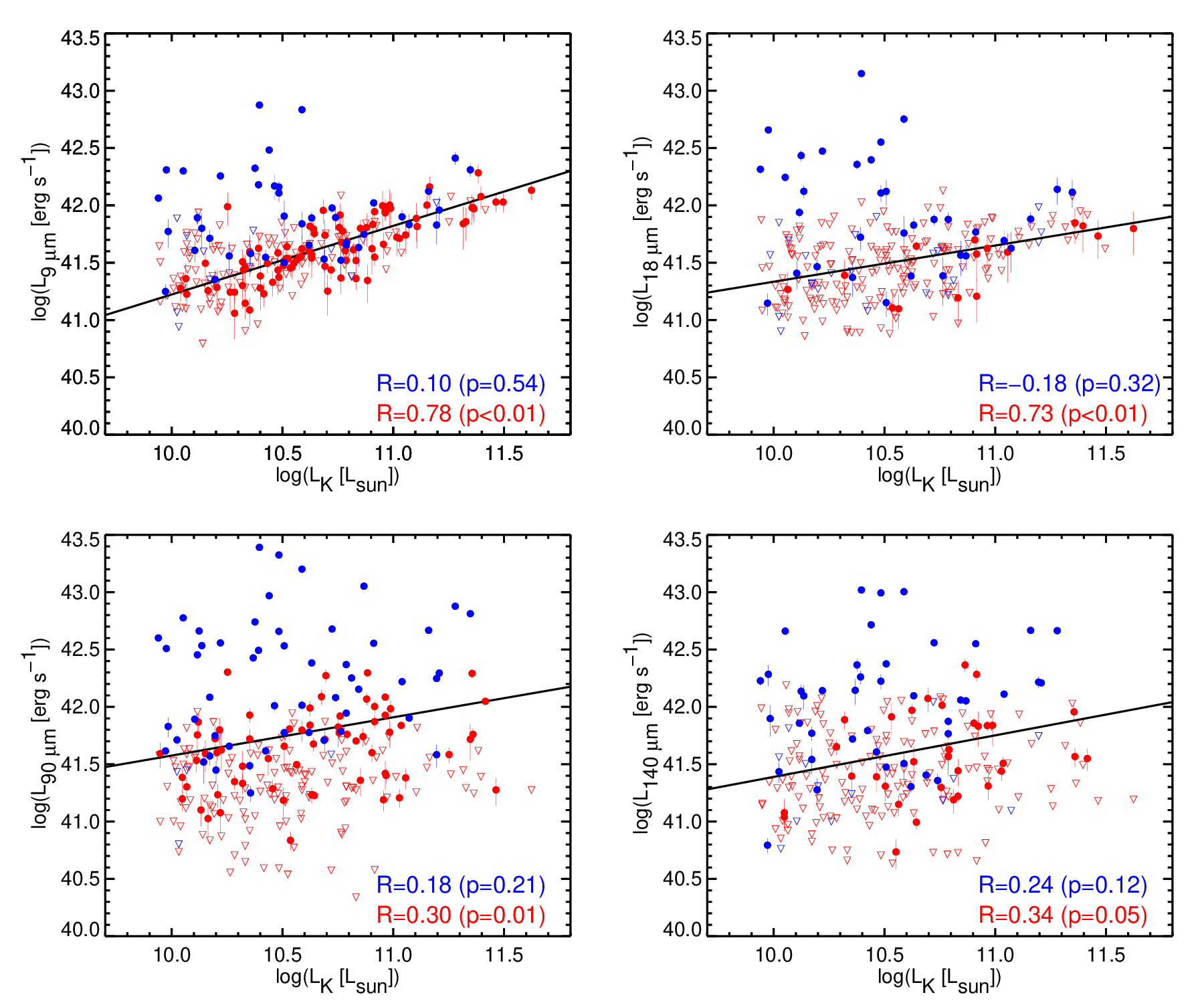}
  \includegraphics[width=11cm]{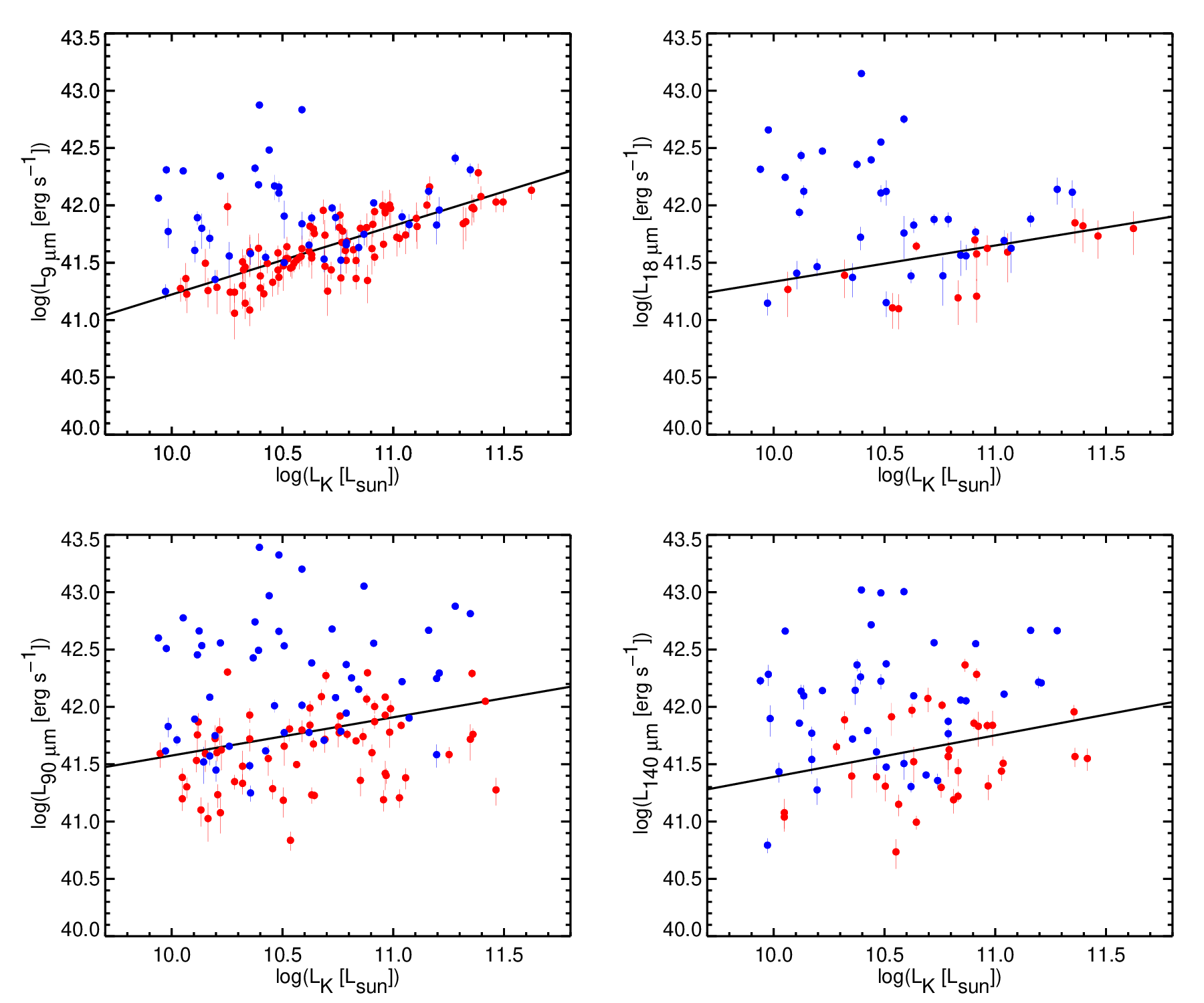}
  \caption{Total $9$, $18$, $90$ and $140$~$\mu$m luminosities versus
    total $K_{\rm s}$-band luminosities, for IR measurements with
    respectively $S/N>2.4$, $2.2$, $2.7$ and $2.8$ (solid
    circles). Conversely, open triangles indicate respectively $2.4$,
    $2.2$, $2.7$ and $2.8\sigma$ upper limits. Blue and red data
    points denote CO-detected and non-CO-detected galaxies,
    respectively. Linear correlation coefficients ($R$) and $p$-values
    for CO-detected and non-CO-detected galaxies are indicated in blue
    and red, respectively, for galaxies above the stated $S/N$
    thresholds. Solid black lines show the best linear fits to the
    non-CO-detected galaxies only, excluding upper limits (see
    Table~\ref{tab:linearfits}). The bottom four panels are the
      same as the top four panels, but without upper limits.}
  \label{fig:ir-k}
\end{figure*}

Figure~\ref{fig:ir-k} shows correlations between the total $9$, $18$,
$90$ and $140$~$\mu$m AKARI luminosities and the total
$K_{\rm s}$-band luminosities listed in \citet{cap11}. Luminosities
are estimated using the galaxy distances listed in \citet{cap11}
throughout this paper. Data points are colour-coded according to
whether the sample galaxies are detected in CO (blue) or not (red) in
\citet{you11}. To quantify the contribution of the stellar emission
from old stellar populations to the AKARI luminosities, we fitted the
correlations of non-CO-detected galaxies with straight lines.  The
best fits are shown in the panels of Fig.~\ref{fig:ir-k} and the
best-fit parameters are listed in Table~\ref{tab:linearfits}. By
subtracting the best fits from the luminosities of CO-detected
galaxies, we can estimate their excess luminosities and assess whether
these correlate with the $K_{\rm s}$-band luminosities. This in turn
establishes if the excess luminosities are of (circum)stellar origin
or not (and thus of interstellar origin).

\begin{table}
  \caption{Parameters of the best linear fits in Fig.~\ref{fig:ir-k}.}
  \label{tab:linearfits}
  \centering
  \begin{tabular}{rcc}
    \hline\hline
    Band & $a$ & $b$ \\
    \hline
    9~$\mu$m & $0.60\pm0.06$ & $35.24\pm0.64$\\
    18~$\mu$m & $0.32\pm0.20$ & $38.18\pm2.12$\\
    90~$\mu$m & $0.33\pm0.04$ & $38.24\pm0.44$\\
    140~$\mu$m & $0.36\pm0.09$ & $37.77\pm1.00$\\
    \hline
  \end{tabular}
  \tablefoot{Data points are fit with ${\rm log}(L_{\rm IR}/{\rm erg}\,{\rm
      s}^{-1})=a\,{\rm log}(L_K/L_\sun)+b$.}
\end{table}

As the sample galaxies cover a rather large range of $K_{\rm s}$-band
luminosities and are spread over that range roughly uniformly in
$\log(L_{K_{\rm s}}/L_\sun)$, it is preferable to do the above fits in
$\log$-$\log$ space (as shown in Fig.~\ref{fig:ir-k}). While we would
ideally use for those fits only robustly detected galaxies (i.e.,
those measurements with $S/N>3$), this would not leave enough data
points for a reliable fit at $18$~$\mu$m. For each band, we therefore
use the lowest $S/N$ threshold yielding only positive measurements
($S/N>2.4$, $2.2$, $2.7$ and $2.8$ at respectively $9$, $18$, $90$ and
$140$~$\mu$m). The linear correlation coefficients ($R$) of those
measurements and the probabilities of obtaining the measured $R$ if
the null hypothesis is true ($p$) are also listed in each panel, to
ascertain the strengths of the correlations. Measurements below those
$S/N$ thresholds are replaced by upper limits of the same significance
in Fig.~\ref{fig:ir-k}.

As can be seen in the top-left panel of Fig.~\ref{fig:ir-k}, the
$9$~$\mu$m luminosities of non-CO-detected galaxies correlate very
well with their $K_{\rm s}$-band luminosities (with a correlation
coefficient $R=0.78$ and $>99\%$ confidence levels). As $K_{\rm s}$
band traces old stellar populations and the $9$~$\mu$m emission of
non-CO-detected galaxies is also expected to be dominated by emission
from the stars themselves, a correlation between these two bands is
expected. However, CO-detected galaxies show excess $9$~$\mu$m
emission, being systematically above the correlation of
non-CO-detected galaxies (or consistent with it), suggesting that
CO-detected galaxies have PAH emission beyond any associated with the
stars. Indeed, the AKARI $9$~$\mu$m band is sensitive to major PAH
bands at $6.2$, $7.7$ and $8.6$~$\mu$m \citep{ish10}. Furthermore, the
excess luminosities of CO-detected galaxies do not correlate well with
their $K_{\rm s}$-band luminosities ($R=-0.14$ and $p=0.36$),
suggesting that their PAHs are not of circumstellar origin but rather
of interstellar origin.

The correlation between the $18$~$\mu$m and $K_{\rm s}$-band
luminosities is also significant for non-CO-detected galaxies
($R=0.73$ and $p<0.01$; top-right panel of Fig.~\ref{fig:ir-k}),
suggesting that the bulk of their $18$~$\mu$m emission is of
circumstellar dust origin. This is supported by the fact that
circumstellar dust generated and heated by old stars is known to be
the dominant emission source in ETGs at $15$ and $24$~$\mu$m
\citep{ath02,tem09}. Most CO-detected galaxies again show a
significant excess above the correlation of non-CO-detected galaxies,
and their excess luminosities is not correlated with their
$K_{\rm s}$-band luminosities ($R=-0.18$ and $p=0.32$). The
$18$~$\mu$m emission of CO-detected galaxies is thus likely due to
interstellar dust heated by relatively young stellar populations or
AGN activity (see \citealt{nyl16} for a study of nuclear radio
emission in ATLAS$^{\rm 3D}$ ETGs). The $9$ and $18$~$\mu$m
luminosities thus strongly suggest that ETGs with detectable molecular
gas also contain significant interstellar dust.

In the bottom two panels of Fig.~\ref{fig:ir-k}, the correlations
between the FIR $90$ and $140$~$\mu$m luminosities (tracing cold dust)
and the NIR $K_{\rm s}$-band luminosities (tracing stellar emission)
of non-CO-detected galaxies are weaker than those in the MIR bands and
are not statistically significant. This appears consistent with the
results of \citet{tem07}, who argued that FIR emission in ETGs
originates from cold dust of interstellar origin, presumably spread
and heated by diffuse starlight and/or hot gas. However, it is worth
noting that the correlations are not totally negligible either
($p=0.01$ and $0.05$ at $90$ and $140$~$\mu$m, respectively). This
raises the interesting possibility that the stellar populations of
ETGs may have cold dust of circumstellar origin, perhaps in shells at
large stellar-centric radii. This should however be verified with
either more accurate or more numerous data. Either way, CO-detected
galaxies again show systematically brighter FIR emission, uncorrelated
with stellar emission, implying that they are even richer in
interstellar cold dust than the norm (i.e., than non-CO-detected
galaxies).

\subsection{PAH and dust emission}
\label{sec:pah-dust}

\begin{figure*}
  \centering
  \includegraphics[width=15cm]{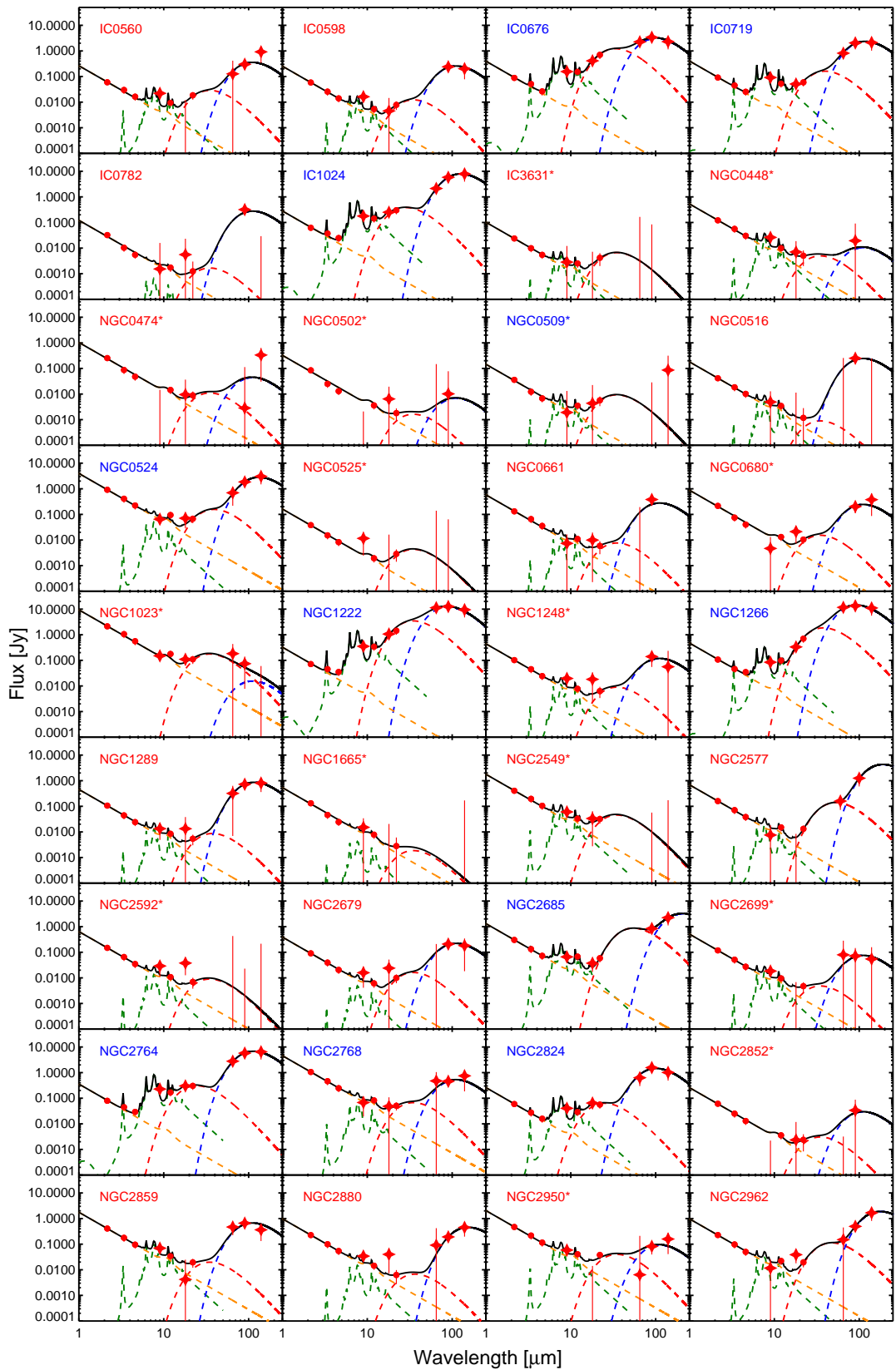}
  \caption{Observed spectral energy distributions (red data points)
    overlaid with the best-fit spectral models (solid black lines),
    for the $231$ successfully fitted ATLAS$^{\rm 3D}$ galaxies.
      AKARI data points are shown as stars, others data points as
      filled circles. Dotted orange, green, red and blue lines
    indicate the best-fit stellar, PAH, warm and cold dust component,
    respectively. The name of the galaxy is indicated in the top-left
    corner of each panel, in blue for CO-detected galaxies and red for
    non-CO-detected galaxies. Galaxies with no AKARI FIR ($65$,
      $90$ or $140$~$\mu$m) detection are marked with an asterisk.}
  \label{fig:seds}
\end{figure*}

\begin{figure*}
  \centering
  \includegraphics[width=\hsize]{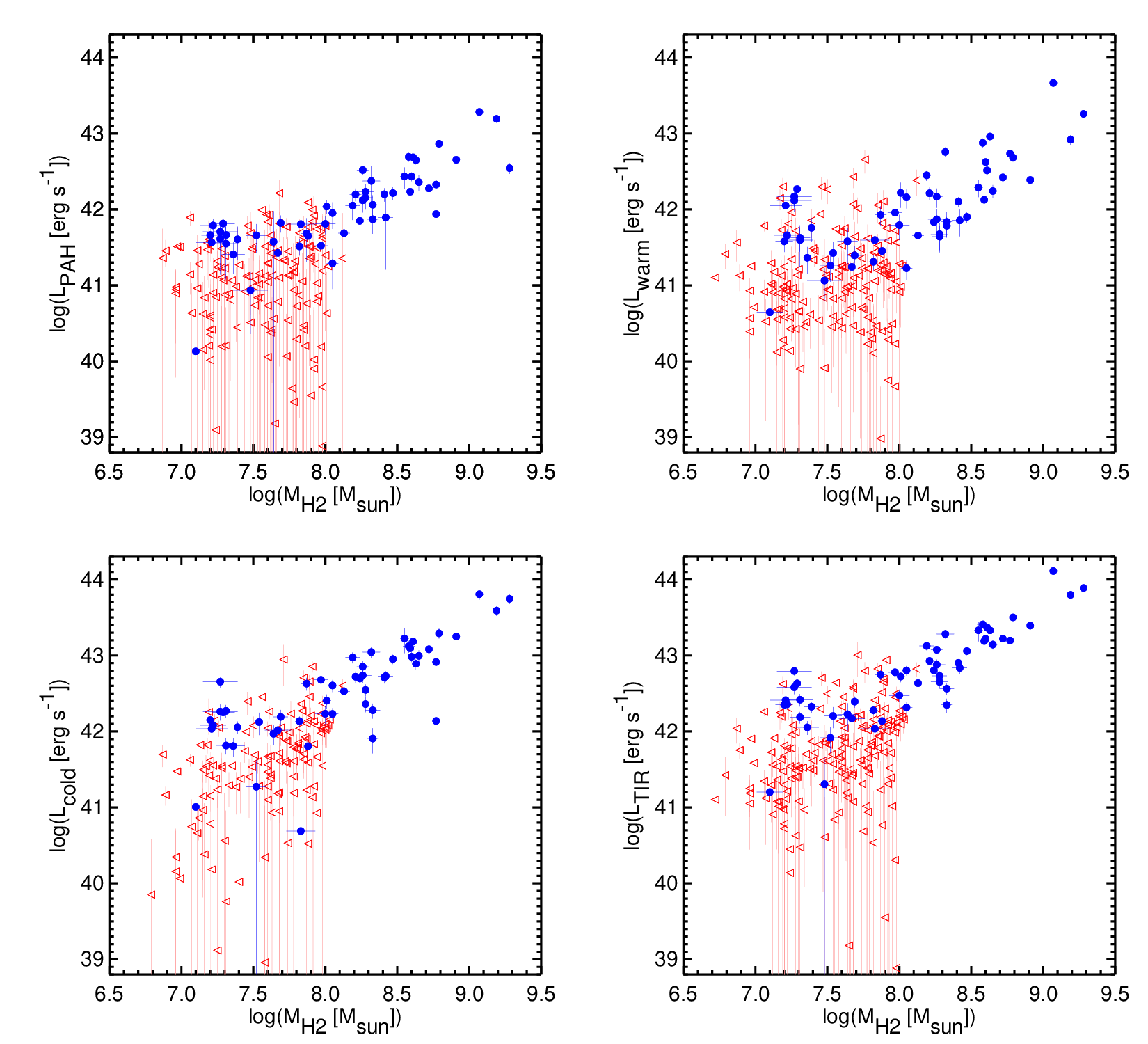}
  \caption{Infrared-derived luminosities (PAH, warm dust, cold dust
    and TIR) versus molecular gas (H$_2$) masses, for the $231$
    successfully fitted ATLAS$^{\rm 3D}$ galaxies. Blue data points
    represent CO-detected galaxies while open red triangles show
    non-CO-detected galaxies (i.e., $M_{{\rm H}_2}$ upper limits).}
  \label{fig:luminosities}
\end{figure*}

To estimate accurate PAH and dust luminosities for our sample
galaxies, we created spectral energy distributions (SEDs) by combining
the 2MASS $K_{\rm s}$-band, WISE and AKARI data. We fitted each SED
with a model composed of emission from stars, PAHs and two dust
components (warm and cold). We described the stellar continuum
emission with a power-law model. In addition, to describe a silicate
feature seen in ETGs around $10$~$\mu$m \citep[e.g.,][]{bre06}, we
added a Gaussian function to the power-law model, where the width and
amplitude of the Gaussian relative to the stellar continuum were
determined from the quiescent elliptical galaxy template of
\citet{kan08}. The \citet{dra07} model was used for the PAH emission,
with a size distribution and ionised fraction typical of the diffuse
ISM of star-forming galaxies (for lack of a better model), while only
the amplitude was allowed to vary. The warm and cold dust components
were each described by a modified blackbody model with emissivity
power-law index $2$ \citep{gal12}, while the amplitude and temperature
were allowed to vary. When fitting the SEDs, we used all
detections regardless of their significance, adopting their errors
as weights ($1/\sigma^2$, where $\sigma$ is the error), with the
additional constraint that the fits must be positive at all
wavelengths.

First, we fitted the SEDs of galaxies robustly detected (i.e.,
$S/N>3$) in two or three FIR bands, by allowing all the parameters to
vary. Then, by fixing the warm and cold dust temperatures to the means
of the best-fit temperatures of those galaxies, we fitted the SEDs of
the remaining galaxies (i.e., galaxies detected in only one or no
FIR band). Although this ignores the differences between random
and systematic uncertainties and is therefore not ideal, we added the
measurement and systematic flux calibration uncertainties in
quadrature when performing the fits.

In our sample, $71$ galaxies are robustly detected in two or three FIR
bands, from which we estimated a mean dust temperature of $83\pm22$~K
for the warm component and $26\pm6$~K for the cold component. The fit
was not acceptable at the $90\%$ confidence level for $3$ of these
objects. For the remaining galaxies, the fit was not acceptable for
$26$ objects. These $29$ objects were thus excluded from the following
analysis. The SEDs and best-fit models of the $231$ remaining
ATLAS$^{\rm 3D}$ ETGs are shown in Fig.~\ref{fig:seds}.

We calculated the PAH and dust luminosities ($L_{\rm PAH}$,
$L_{\rm warm}$ and $L_{\rm cold}$) by integrating the aforementioned
best-fits of each component over the wavelength range
$5$\,--\,$1000$~$\mu$m. The luminosities thus derived are listed in
Table~\ref{tab:luminosities}, and the total IR (TIR) luminosities
($L_{\rm TIR}$) were estimated by simply taking the sums of those
three components. To check the reliability of our PAH
luminosities, we compared them to the PAH luminosities derived by
\citet{sha10} using the non-stellar Spitzer $8$~$\mu$m emission (as
done in Sect.~\ref{sec:akari} for our $90$ and $140$~$\mu$m flux
densities). The PAH luminosities are almost all consistent within
the uncertainties, and are proportional with a slope of
approximately unity and a small intrinsic scatter of 0.2 dex.
Figure~\ref{fig:luminosities} shows $L_{\rm PAH}$, $L_{\rm warm}$,
$L_{\rm cold}$ and $L_{\rm TIR}$ plotted against H$_2$ mass, the
latter calculated from the integrated CO line intensity assuming a
Galactic conversion factor \citep[see][]{you11}. It is clear that the
IR luminosities of CO-detected galaxies (here stellar
emission-subtracted by construction) are well correlated with their
molecular gas masses, thus confirming a tight connection between dust
and molecular gas even in ETGs \citep[see also][]{com07}. Among the IR
luminosities, $L_{\rm warm}$ has a relatively large scatter,
suggesting that dust heating by young stellar populations and/or AGN
can vary from galaxy to galaxy.

\subsection{Star formation rates}
\label{sec:sfrs}

\begin{figure*}
  \centering
  \includegraphics[width=\hsize]{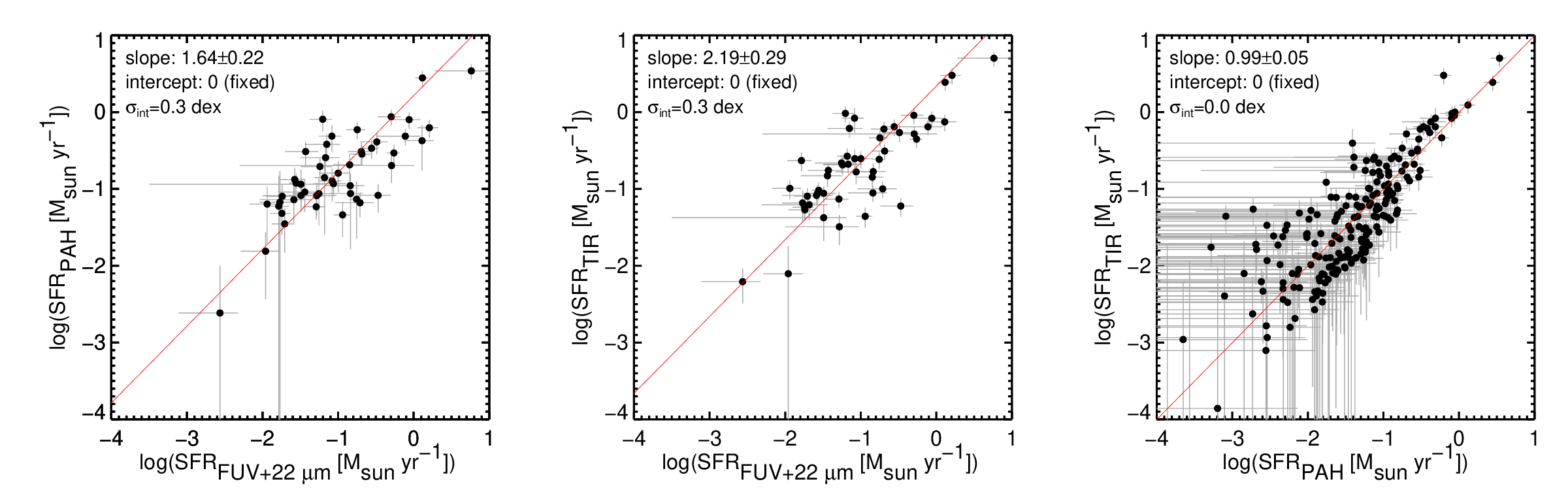}
  \caption{Comparisons of our PAH- and TIR-derived star formation
    rates and the FUV+$22$~$\mu$m star formation rates of
    \citet{dav14}, the latter for CO-detected galaxies only.
      Solid red lines show the best-fit linear relations with nul
      intercepts, while the labels list the best-fit slopes, fixed
      intercepts and intrinsic scatters around the best-fit
      relations.}
  \label{fig:sfrs_fuv}
\end{figure*}

\begin{figure*}
  \centering
  \includegraphics[width=\hsize]{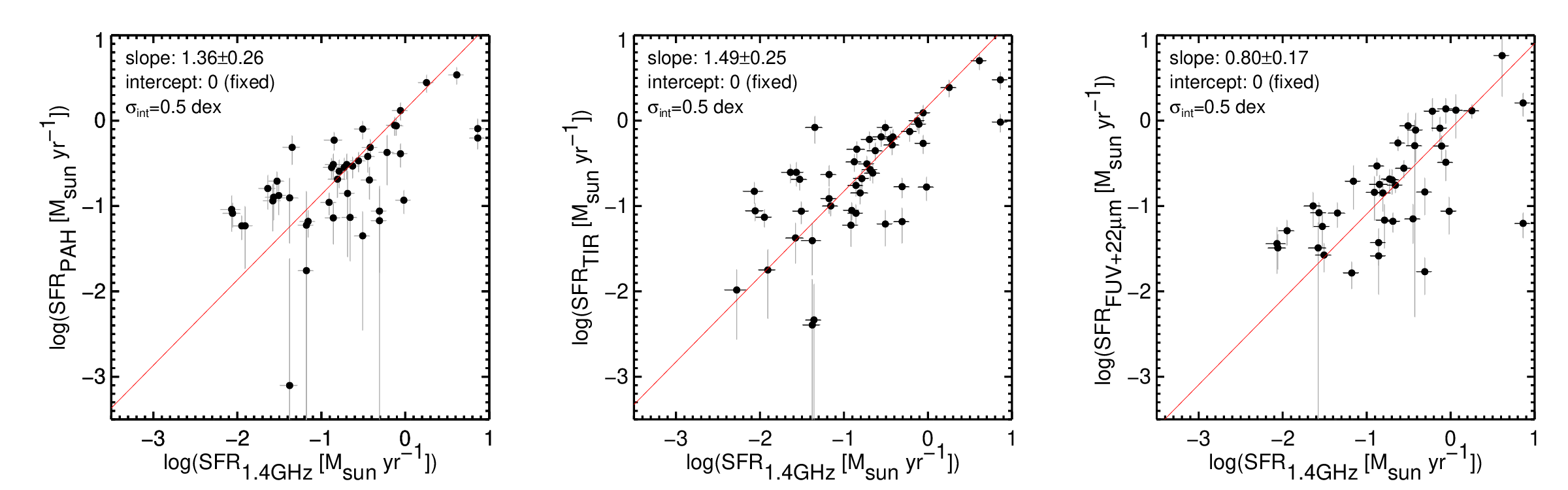}
  \caption{Comparisons of our PAH- and TIR-derived star formation
      rates and the FUV+$22$~$\mu$m star formation rates of
      \citet{dav14} with 1.4~GHz star formation rates derived from the
      measurements of \citet{nyl17}, for all galaxies in common. Solid
      red lines show the best-fit linear relations with nul
      intercepts, while the labels list the best-fit slopes, fixed
      intercepts and intrinsic scatters around the best-fit
      relations.}
  \label{fig:sfrs_radio}
\end{figure*}

To estimate the current SFRs of the $231$ ATLAS$^{\rm 3D}$ ETGs with
good SED fits, we utilised the PAH and TIR luminosities calculated
from those fits. In particular, ionised PAHs are known to be good
tracers of star formation, as they are exposed to the radiation fields
of photo-dissociation regions around relatively young stars, and the
AKARI $9$~$\mu$m band is sensitive to the $6.2$, $7.7$ and
$8.6$~$\mu$m features of ionised PAHs \citep{ish10}. We note that
relatively quiescent ETGs tend to show unusually high PAH
$11.3$~$\mu$m/PAH $7.7$~$\mu$m ratios, indicating a dominance of
neutral PAHs not related to star formation
\citep[e.g.,][]{kan05,kan08,pan11}, but our results are not likely to
be sensitive to this unusual interband ratio as $L_{\rm PAH}$ is
mostly determined from the AKARI $9$~$\mu$m band that excludes the
$11.3$~$\mu$m feature. Furthermore, the WISE $12$~$\mu$m-band
fluxes do not indicate the presence of unusually strong PAH
$11.3$~$\mu$m features. To convert $L_{\rm PAH}$ to a SFR, we
adopted the method described in \citet{shi16}, in which the sum of the
$6.2$, $7.7$ and $11.3$~$\mu$m features is used to estimate the
SFR. The ratio of these PAH bands to the total PAH emission is $0.65$
in the PAH template adopted for our SED fits (a value typical of
star-forming galaxies; \citealt{smi07}), so we simply propagate this
fraction in our calculations. To convert $L_{\rm TIR}$ to a SFR, we
used the conversion of \citet{ken98} with a correction to the Kroupa
initial mass function \citep{hao11,mur11}.

For the majority of our $231$ ETGs, the current SFR is estimated to be
in the range $0.01$\,--\,$1$~$M_{\sun}$~yr$^{-1}$. Furthermore, as
shown in the right panel of Fig.~\ref{fig:sfrs_fuv}, the SFRs
estimated from PAH luminosities are in good agreement with those
estimated from TIR luminosities, with an average ratio of
SFR$_{\rm PAH}$/SFR$_{\rm TIR}=1.0\pm0.1$ and nul intrinsic scatter
(given the large uncertainties; estimated again as in Sect.~\ref{sec:akari}).

For $56$ CO-detected ETGs of the ATLAS$^{\rm 3D}$ sample,
\citet{dav14} estimated the current SFR with a combination of FUV and
$22$~$\mu$m emission. In the left and middle panels of
Fig.~\ref{fig:sfrs_fuv}, we therefore compare those SFRs with the PAH
and TIR SFRs calculated here. Our SFRs are generally in good
agreements with those of \citet{dav14}, but they are
systematically larger by a factor of $\approx2$ at SFRs smaller than
${\approx}0.1$~$M_{\sun}$~yr$^{-1}$ (SFR$_{\rm 
PAH}$/SFR$_{{\rm FUV}+22 \mu{\rm m}}=1.6\pm0.2$ and
SFR$_{\rm TIR}$/SFR$_{{\rm FUV}+22 \mu{\rm m}}=2.2\pm0.3$, with
intrinsic scatters of $0.3$~dex, for all galaxies in common). This
systematic difference is presumably due to the specific physical
conditions in the star-formation regions of ETGs, that will be
discussed in Sect.~\ref{sec:sfprops}.

Low-frequency ($1.4$~GHz) radio continuum observations of $72$
ATLAS$^{\rm 3D}$ ETGs were presented by \citet{nyl17}. We therefore
estimated SFRs from these measurements using the conversion of
\citet{mur11}, and we compare these estimates with our PAH- and
TIR-derived estimates in Fig.~\ref{fig:sfrs_radio}. As shown in the
left and middle panels, our measurements are generally in good
agreements with those derived from \citet{nyl17}, but they are
systematically larger by factors of SFR$_{\rm
 PAH}$/SFR$_{\rm 1.4GHz}=1.4\pm0.3$ and SFR$_{\rm
 TIR}$/SFR$_{\rm 1.4GHz}=1.5\pm0.3$, with intrinsic scatters of
$0.5$~dex (for all galaxies in common). As can be expected from the
discussion in the paragraph above, the $1.4$~GHz- and
FUV+$22$~$\mu$m-derived SFRs are then in rough agreement
(SFR$_{{\rm FUV}+22 \mu{\rm m}}$/SFR$_{\rm 1.4GHz}=0.8\pm0.2$, with
an intrinsic scatter of $0.5$~dex for galaxies in common; see the
right panel of Fig.~\ref{fig:sfrs_radio}). These results will also
be discussed in Sect.~\ref{sec:sfprops}.


\section{Discussion}
\label{sec:discussion}

\subsection{Relations between SFRs and gas masses}
\label{sec:sfr-mgas}

\begin{figure*}
  \centering
  \includegraphics[width=\hsize]{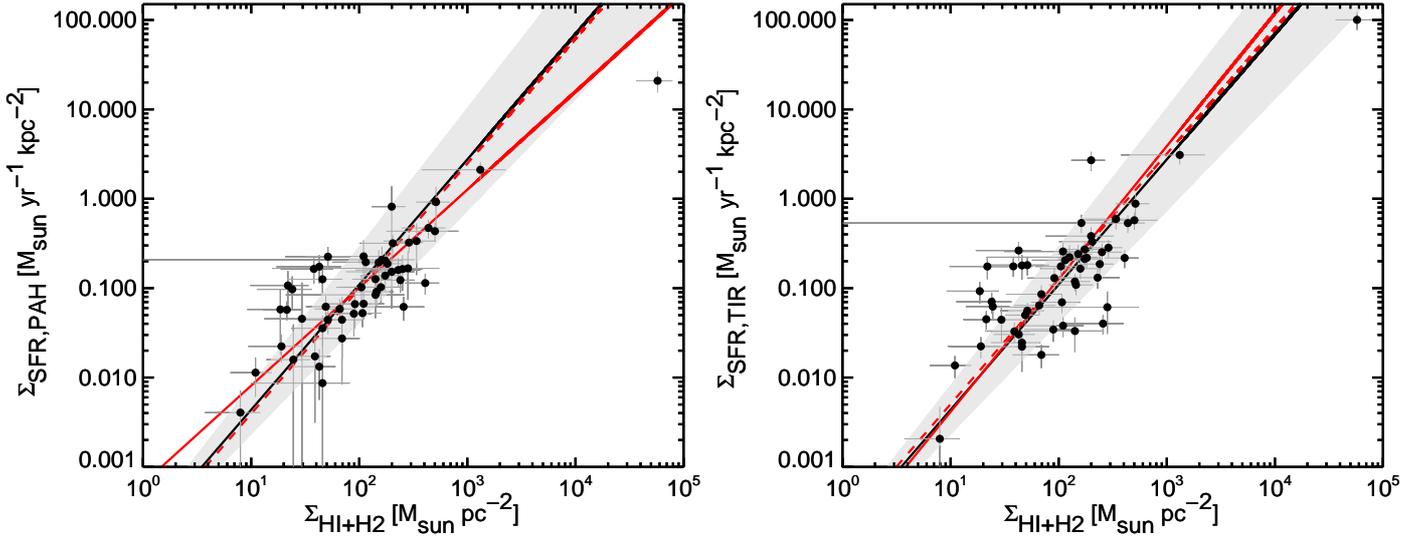}
  \caption{Star formation rates (PAH- and TIR-derived) versus total
    gas surface densities (i.e., KS law) for CO-detected
    galaxies. Solid black lines and shaded grey regions show
    respectively the \citet{ken98} relation for local star-forming
    galaxies (with power-law index $n=1.4$) and its intrinsic
    scatter. Solid and dashed red lines show the best linear fits with
    respectively free and fixed ($n=1.4$) slope. The galaxy
      located in the top-right of each panel is NGC~1266, known to
      have strongly suppressed star formation. It was excluded from
      the fits.}
  \label{fig:ks}
\end{figure*}

\begin{figure*}
  \centering
  \includegraphics[width=\hsize]{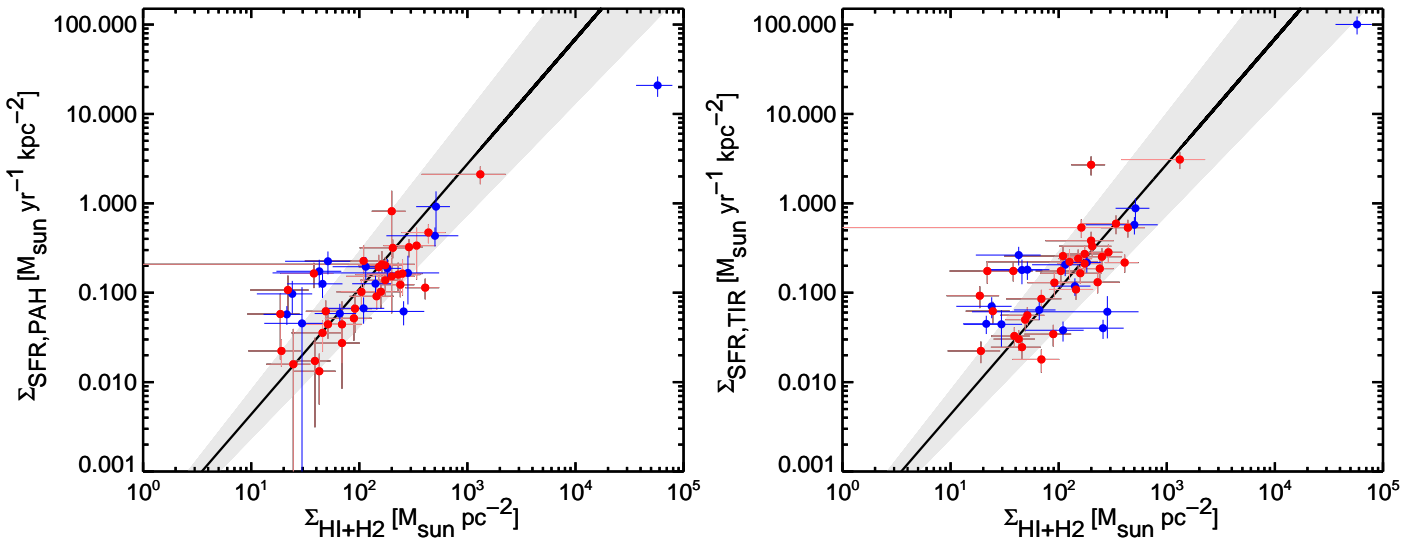}
  \caption{Same as Fig.~\ref{fig:ks}, but with galaxies with
    kinematically-aligned and misaligned stars and ionised gas shown
    in red and blue, respectively.}
  \label{fig:ks-align}
\end{figure*}

Star formation in galaxies is characterized by a power-law relation
between the current SFR ($\Sigma_{\rm SFR}$) and cold gas mass
($\Sigma_{\rm gas}$) surface densities, known as the Kennicutt-Schmidt
(KS) law ($\Sigma_{\rm SFR}\,\propto\,\Sigma_{\rm gas}^n$), where the
power-law index $n$ is found to be ${\approx}1.4$ for local
star-forming galaxies \citep{ken98} but varies from $1$ to $2$ from
galaxy to galaxy, reflecting differences in the star formation
processes \citep[e.g.,][]{tan98,mis06}. The power-law index is also
generally closer to unity for molecules tracing gas denser than CO
(e.g.,\ HCN; \citealt{gao04}). \citet{dav14} investigated the KS law
of $56$ CO-detected ATLAS$^{\rm 3D}$ galaxies using their SFRs
estimated from FUV and $22$~$\mu$m emission. Here, we revisit the KS
law for the same $56$ ETGs using our SFRs estimated from PAH and TIR
emission.

To calculate $\Sigma_{\rm SFR}$ and $\Sigma_{\rm gas}$, we use our own
SFRs but the gas masses and source sizes listed in \citet{dav14},
where the gas masses were derived by summing the CO and central
\ion{H}{I} gas masses, and the sizes were measured using
interferometric CO, spatially-resolved MIR, optical or FUV
observations. The resulting $\Sigma_{\rm SFR}$\,--\,$\Sigma_{\rm gas}$
correlations are shown in Fig.~\ref{fig:ks}. The solid black lines and
shaded grey regions show respectively the KS law of local star-forming
galaxies with $n=1.4$ and its intrinsic scatter \citep{ken98}. The
solid and dashed red lines show the best linear fits to the data with
respectively free and fixed ($n=1.4$) slope. The best-fit parameters
are listed in Table~\ref{tab:ks}. We note that the galaxy
NGC~1266, that is systematically located in the top-right of each
panel of Fig.~\ref{fig:ks}, is known to have strongly suppressed
star formation and thus to be an outlier in the star-formation
relation \citep[see][]{ala11,ala15,nyl13}. We have thus excluded it
from the fits. Figure~\ref{fig:ks} shows that the majority of
CO-detected (and thus potentially star-forming) ETGs do follow the KS
law of local star-forming galaxies. In particular, the fixed-slope
fits have no offset with respect to the \citet{ken98} relation,
suggesting that the SFEs of local ETGs are similar to those of local
star-forming galaxies. This is contrary to the results of
\citet{dav14}, that suggested smaller SFEs for the same sample
galaxies, by a factor of $2$\,--\,$3$. This difference can be directly
attributed to the similarly higher SFRs calculated here from the AKARI
data (see Sect.~\ref{sec:sfrs} and
Figs.~\ref{fig:sfrs_fuv} and \ref{fig:sfrs_radio}). We come back to this
difference in Sect.~\ref{sec:sfprops}.

\begin{table}
  \caption{Parameters of the best linear fits in Fig.~\ref{fig:ks}.}
  \label{tab:ks}
  \centering
  \begin{tabular}{lcc}
    \hline\hline
    SFR tracer & $n$ & $c$ \\
    \hline
    PAH & $1.10\pm0.11$ & $-3.19\pm0.26$\\
    PAH & $1.40$ (fixed) & $-3.81\pm0.04$\\
    Total IR & $1.49\pm0.13$ & $-3.87\pm0.27$\\
    Total IR & $1.40$ (fixed) & $-3.70\pm0.03$\\
    \hline
  \end{tabular}
  \tablefoot{Data points are fit with $\log(\Sigma_{\rm
      SFR}/M_\sun\,{\rm pc}^{-2})=n\,\log(\Sigma_{\rm
      gas}/M_\sun\,{\rm pc}^{-2})+c$.} 
\end{table}

When left free, the slopes of the power-law fits are $n=1.10\pm0.11$
and $n=1.49\pm0.13$ for the PAH and TIR SFRs, respectively. The
significant deviation of the PAH SFR slope from $n=1.4$ appears to
be caused by several galaxies with relatively high $\Sigma_{\rm SFR}$ at
$\Sigma_{\rm gas}\lesssim0.6\times10^2$~$M_\sun$~pc$^{-2}$.  Those
galaxies may have a different star-formation mode (e.g., widespread
versus circumnuclear) or a different ISM-acquisition mode (e.g.,
internal versus external). The origin of the gas can be constrained by
examining the kinematic misalignment between stars and
gas. Kinematically-misaligned gas likely indicates an external origin
(e.g.,\ minor merger or accretion from filaments), while
kinematically-aligned gas is expected from both internal (e.g.,\
stellar mass loss or remnant from the galaxy formation event) and
occasionally external gas. Figure~\ref{fig:ks-align} is analogous to
Fig.~\ref{fig:ks}, but we colour code the galaxies according to their
star-ionised gas kinematic (mis)alignment, when such a measurement is
available. \citet{dav11} measured star-gas kinematic (mis)alignments
for $35$ CO-detected galaxies mapped interferometrically in CO, but
$51$ CO-detected galaxies mapped with optical integral-field
spectroscopy in ionised gas, and they showed that both measurements
are always in agreement. We therefore use the latter measurements
here. Interestingly, Fig.~\ref{fig:ks-align} reveals that many of the
galaxies with relatively high $\Sigma_{\rm SFR}$ at
$\Sigma_{\rm gas}\lesssim0.6\times10^2$~$M_\sun$~pc$^{-2}$ (and none
of the galaxies with relatively low $\Sigma_{\rm SFR}$ at those same
$\Sigma_{\rm gas}$) have kinematically-misaligned gas, indicating that
they have acquired their ISM from external sources. Galaxies with
kinematically-misaligned gas also seem to have higher scatters
around the best-fit relations than galaxies with
kinematically-aligned gas, perhaps suggesting a more bursty and thus
temporally variable recent star formation history.

\subsection{Relations between SFRs and stellar masses}
\label{sec:sfr-mstar}

\begin{table*}
  \caption{Parameters of the best linear fits in Fig.~\ref{fig:sfms}.}
  \label{tab:sfms}
  \centering
  \begin{tabular}{lccc}
    \hline\hline
    SFR tracer & Mass-weighted age (Gyr) & $a$ & $b$ \\
    \hline
    PAH & $\phantom{1}3$\,--\,$\phantom{1}7$ & $0.93\pm0.23$ & $-10.12\pm2.33$\\
    PAH & $\phantom{1}7$\,--\,$11$ & $0.72\pm0.20$ & \phantom{1}$-8.32\pm2.10$\\
    PAH & $11$\,--\,$14$ & $1.09\pm0.22$ & $-12.94\pm2.43$\\
    Total IR & $\phantom{1}3$\,--\,$\phantom{1}7$ & $0.80\pm0.17$ & \phantom{1}$-8.82\pm1.71$\\
    Total IR & $\phantom{1}7$\,--\,$11$ & $0.76\pm0.13$ & \phantom{1}$-8.95\pm1.37$\\
    Total IR & $11$\,--\,$14$ & $0.84\pm0.09$ & $-10.32\pm1.04$\\
    \hline
  \end{tabular}
  \tablefoot{Data points are fit with $\log({\rm SFR}/M_\sun\,{\rm
      yr}^{-1})=a\,\log(M_\star/M_\sun)+b$.}
\end{table*}

\begin{figure*}
  \centering
  \includegraphics[width=\hsize]{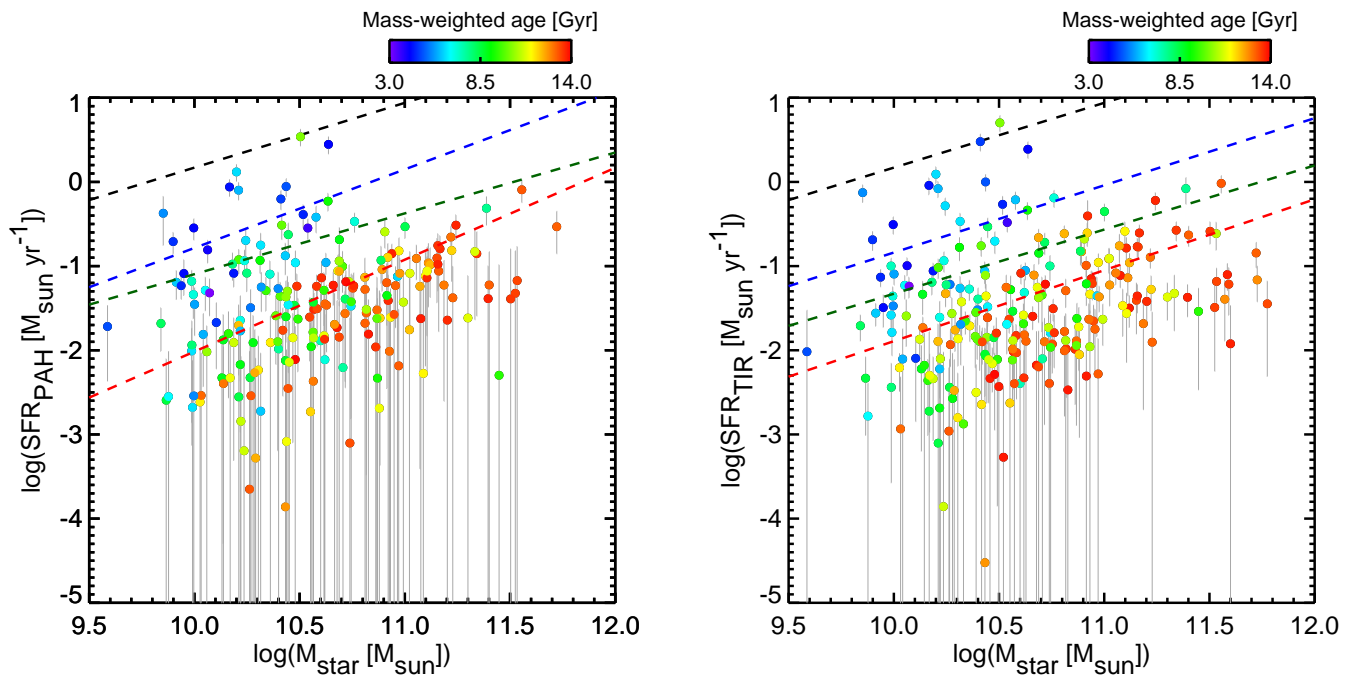}
  \caption{Correlations between the star formation rates (PAH- and
    TIR-derived) and stellar masses of the $231$ successfully fitted
    ATLAS$^{\rm 3D}$ galaxies. Data points are colour-coded according
    to their mass-weighted stellar population age
    \citep{mcd15}. Dotted black lines show the star formation main
    sequence of local star-forming galaxies \citep{elb07}. Dotted
    blue, green and red lines show the best linear fits for the age
    range $3$\,--\,$7$, $7$\,--\,$11$ and $11$\,--\,$14$~Gyr,
    respectively.}
  \label{fig:sfms}
\end{figure*}

\begin{figure*}
  \centering
  \includegraphics[width=\hsize]{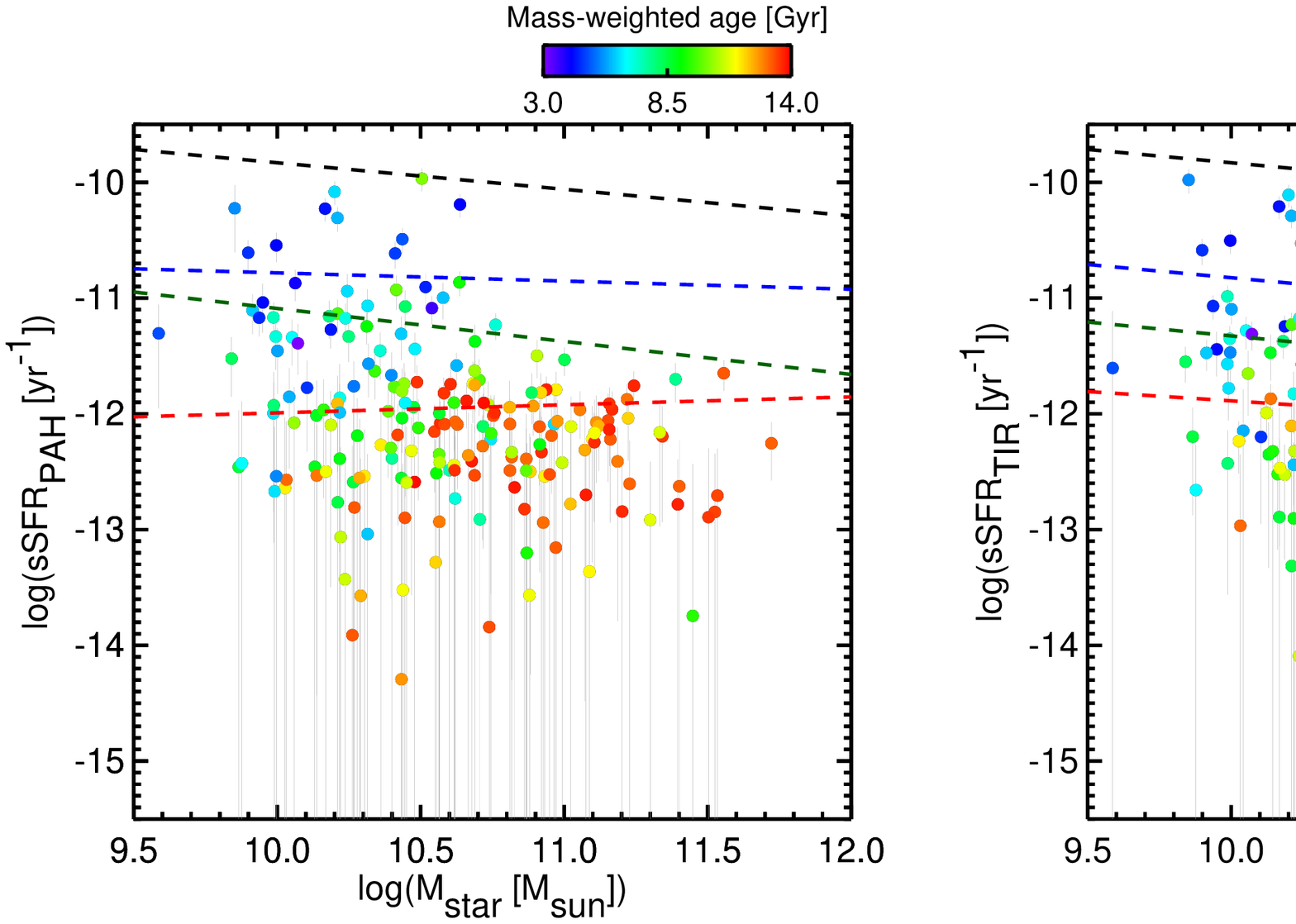}
  \caption{Same as Fig.~\ref{fig:sfms}, but for the specific star
    formation rates.}
  \label{fig:ssfr}
\end{figure*}

In Fig.~\ref{fig:sfms}, the PAH and TIR SFRs of our sample galaxies
are plotted against their stellar masses $M_\star$, the latter
calculated from optical photometry and stellar mass-to-light ratios
measured from dynamical models \citep{cap13}. The data points are also
colour-coded according to the mass-weighted stellar population age of
each galaxy, obtained by fitting stellar population synthesis models
to the optical spectrum of the inner effective radius
\citep{mcd15}. We use mass-weighted ages rather than the more common
luminosity-weighted ages as the former better represent the age of the
bulk of the stellar populations, being less affected by a sprinkling
of young stars onto a dominant old stellar population. The dotted
black lines show the so-called star-formation main sequence, i.e., the
correlation between current SFR and stellar mass of local star-forming
galaxies \citep{elb07}. The dotted blue, green and red lines are
linear fits to our ETG galaxies in the age range $3$\,--\,$7$,
$7$\,--\,$11$ and $11$\,--\,$14$~Gyr, respectively. The best-fit
parameters are listed in Table~\ref{tab:sfms}. A similar figure for
the SAURON sample of $48$ ETGs was presented by \citet{sha10}, but the
much larger ATLAS$^{\rm 3D}$ sample of $260$ ETGs studied here ($231$
with relevant measurements) offers much clearer insight into the
relationship between current SFR, stellar mass and stellar population
age.

Figure~\ref{fig:sfms} shows that, at fixed stellar mass, all our
sample ETGs have a lower current SFR than that of local star-forming
galaxies, confirming that local ETGs are quiescent
galaxies. Figure~\ref{fig:sfms} also shows that, overall (and at least
in the recent history of the universe), the SFRs of ETGs must decrease
as they evolve (i.e., ETGs move to the lower-right of the panels as
they become older and more massive). However, while at fixed mass the
current SFRs decrease strongly with age (here by about $1.5$~dex), at
fixed stellar age the current SFRs increase with stellar mass. In
fact, at fixed stellar age the SFRs and stellar masses of our ETGs are
correlated similarly to those of local star-forming galaxies, with
power-law slopes near unity (see Table~\ref{tab:sfms}).  This is shown
more clearly in Fig.~\ref{fig:ssfr}, that shows the specific SFRs
(sSFRs, i.e., ${\rm SFR}/M_\star$) of our sample galaxies as a
function of stellar mass. Indeed, while the current sSFRs of our ETGs
strongly decrease with age as expected from Fig.~\ref{fig:sfms}, they
are nearly independent of stellar mass at any age. Similar trends are
reported for galaxies with low SFRs by \citet{sai16}, who studied both
early- and late-type galaxies (see their Figs.~2 and 4). This
suggests that local ETGs acquire their cold ISM necessary to form
stars by mechanisms related to stellar mass, as do normal star-forming
galaxies \citep[e.g.,][]{dave11}.

We note here that the sSFRs and stellar ages we discuss are
independent of each other and are not correlated by construction. As
our SFRs (estimated from MIR photometry) trace only young stars
while our total stellar masses (estimated from optical photometry
and stellar dynamical modelling) include both young and old stars,
our sSFRs trace the ratio of young to both young and old
stars. On the other hand, our stellar ages (estimated from optical
spectroscopy) trace the ratio of young to old stars only. Our SFRs
are also (nearly) instantaneous measures (tracing stars of ages
typically $<100$~Myrs), while our stellar ages are also affected by
(and nearly always degenerate with) intermediate age stars
($\ge1$~Gyr).

\subsection{Star formation in ETGs}
\label{sec:sfprops}

\begin{figure*}
  \centering
  \includegraphics[width=\hsize]{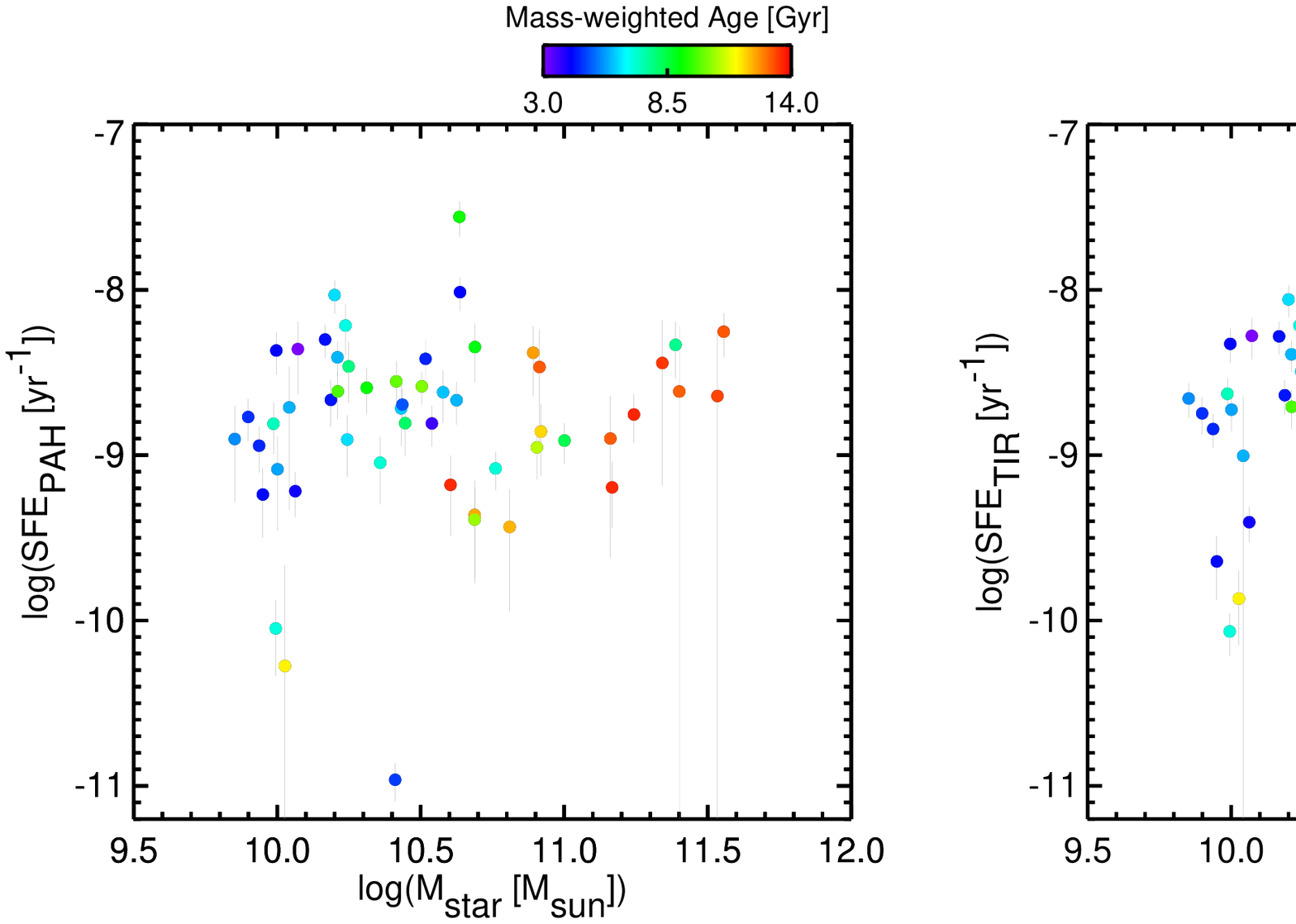}
  \caption{Star formation efficiencies (PAH- and TIR-derived) versus
    stellar masses for CO-detected galaxies. Data points are
    colour-coded according to their mass-weighted stellar population
    age \citep{mcd15}, as in Fig.~\ref{fig:sfms}.}
  \label{fig:sfes}
\end{figure*}

Based on current SFRs estimated from FUV and $22$~$\mu$m emission,
\citet{dav14} argued that local ETGs have SFEs lower than those of
local star-forming galaxies, an effect they assigned to the particular
dynamical properties of ETGs, such as increased central disc stability
and/or galactic shear in the central regions where molecular gas is
found. However, our results show that the SFEs of (the same) local
ETGs estimated from AKARI PAH and TIR emission are similar to those of
local star-forming galaxies (see Figs.~\ref{fig:sfrs_fuv} and
\ref{fig:ks}), in agreement with the smaller study of \citet{sha10}
utilizing Spitzer PAH emission. As TIR emission can be enhanced by
low-luminosity AGN and old stellar populations, that are unrelated to
star-formation, the discrepancy of the TIR-estimated SFRs is likely
easily explained. The discrepancy of the PAH-estimated SFRs is however
not so easily explained, and may be due to the fact that PAH emission
(like TIR emission) can trace star-formation regions with softer UV
radiation fields than those traced in the FUV and at $22$~$\mu$m
\citep[e.g.,][]{com11}. The higher PAH SFRs therefore apparently call
for less massive star formation. Indeed, it is suggested that galaxies
with low SFRs preferentially form low-mass stars (rather than
high-mass stars; e.g., \citealt{hov08,gun11}), which would result in
softer radiation fields in ETGs. The ratios of the [\ion{C}{ii}]
$158$~$\mu$m line to FIR fluxes also support radiation fields softer
in ETGs than in spiral galaxies \citep{mal00}. While old stars in ETGs
can and do emit in the UV (UV-upturn phenomenon; \citealt{yi97}),
\citet{dav14} argued that this does not significantly affect the
FUV-based SFRs of the ATLAS$^{\rm 3D}$ ETGs. Interestingly, this
preference for low-mass stars is consistent with a flurry of recent
works arguing that ETGs have bottom-heavy initial mass functions
\citep[IMFs; see, e.g.,][]{dok10, cap12}.

Of course, just like the FUV and $22$~$\mu$m estimates, the
$1.4$~GHz SFR estimates derived from the \citeauthor{nyl17}'s
(\citeyear{nyl17}) radio continuum measurements are also
systematically lower than the PAH- and TIR-derived estimates (see
Fig.~\ref{fig:sfrs_radio}). As the $1.4$~GHz continuum emission is
due to non-thermal synchrotron emission associated with supernova
remnants (and thus ultimately the explosions of massive stars; e.g.,
\citealt{mur11}), this trend is also consistent with a scenario
whereby massive star formation is inefficient in ETGs.

Figure~\ref{fig:sfes} shows the current SFEs of CO-detected ETGs as a
function of their stellar masses, with the data points colour-coded as
a function of mass-weighted stellar age. No dependence is apparent as
a function of either stellar mass or age. Combined with the continuous
decrease of the sSFR\,--\,$M_{\rm star}$ relations with stellar age
seen in Fig.~\ref{fig:ssfr}, this implies that the current cold gas
fraction ($M_{\rm gas}/M_\star$) of ETGs must decrease with stellar
age (at fixed stellar mass;
${\rm sSFR}\equiv{\rm SFR}/M_\star={\rm SFE}\times M_{\rm
  gas}/M_\star$). We show this trend in Fig.~\ref{fig:gas}, where
$M_{\rm gas}/M_\star$ is plotted as a function of mass-weighted
stellar age. The same trend is reported for nearby massive galaxies
($\log(M_\star/M_\sun)>10$) by \citet{sai11,sai16}, although our study
extends the trend to lower gas fractions
($M_{\rm gas}/M_\star{\approx}10^{-4}$). Hence our results suggest
that the decrease of the current SFR in local ETGs is likely due to a
decrease of their cold gas fractions, rather than a suppression of
star formation (via, e.g., AGN feedback or specific dynamical
properties).

\begin{figure}
  \centering
  \includegraphics[width=\hsize]{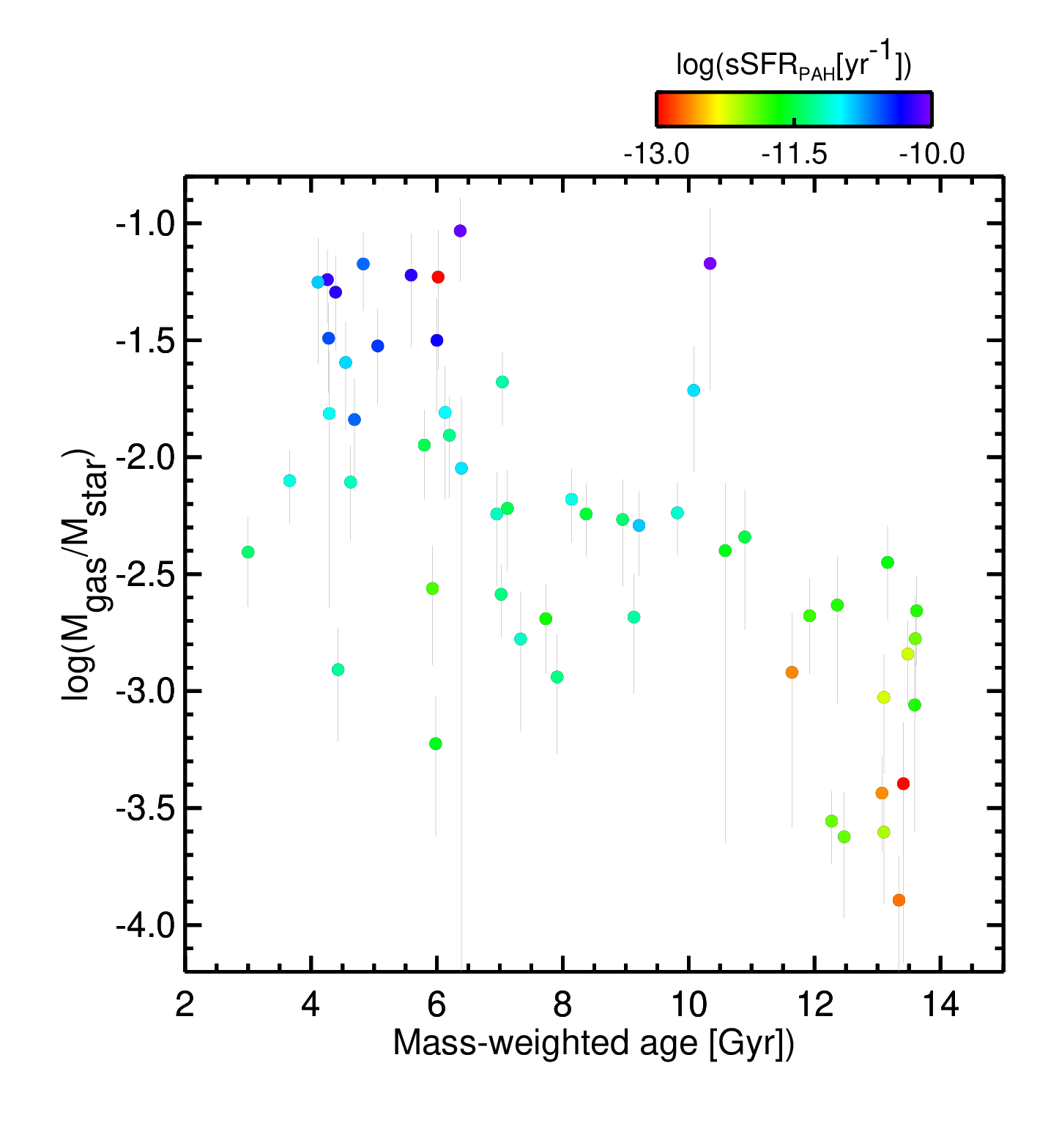}
  \caption{Cold gas fraction versus mass-weighted stellar population
    age \citep{mcd15} for CO-detected galaxies. Data points are
    colour-coded according to their PAH-derived specific star
    formation rates.}
  \label{fig:gas}
\end{figure}


\section{Conclusions}
\label{sec:conclusions}

We have conducted a systematic study of star formation in the $260$
local ETGs of the ATLAS$^{\rm 3D}$ survey, using newly measured IR
fluxes from the AKARI all-sky diffuse maps combined with WISE and
2MASS literature data. In the AKARI FIR bands, $117$ ($45\%$)
galaxies are detected in at least one band, while $71$ ($27\%$)
galaxies are detected in at least two bands. The $9$ and
$18$~$\mu$m luminosities of the non-CO-detected galaxies are
correlated with the stellar luminosities, showing that they trace
respectively stellar and circumstellar dust emission. On the other
hand, CO-detected galaxies show an excess above these correlations,
uncorrelated with the stellar luminosities, indicating that they
likely contain PAHs and dust of interstellar origin. There are at best
weak correlations between the $90$ and $140$~$\mu$m luminosities and
the stellar luminosities of non-CO-detected galaxies (but again an
excess for CO-detected galaxies), suggesting that FIR emission in ETGs
primarily originates from cold dust unrelated to the stars.

We estimated the IR luminosities ($L_{\rm PAH}$, $L_{\rm warm}$,
$L_{\rm cold}$ and $L_{\rm TIR}$) of the sample ETGs by decomposing
their SEDs with a dust model. For the CO-detected galaxies, all of the
IR luminosities correlate well with the H$_2$ masses, confirming a
tight connection between interstellar dust and molecular gas in local
ETGs. Utilizing $L_{\rm PAH}$ and $L_{\rm TIR}$ as current SFR
indicators, we measure typical current SFRs of
$0.01$\,--\,$1$~$M_\sun$~yr$^{-1}$. These SFRs are generally in good
agreement with previously measured SFRs using FUV and $22$~$\mu$m
emission as well as $1.4$~GHz radio continuum emission, but they
are systematically higher by a factor of $\approx1.5$\,--\,$2$,
particularly at low SFRs. This discrepancy may be due to the fact that
PAH and TIR emission can trace star-formation regions with softer
radiation fields than those traced in the FUV, $22$~$\mu$m and
$1.4$~GHz bands. The majority of ETGs appear to follow the standard
KS law of local star-forming galaxies, with no offset, indicating that
they have similar current SFEs. This is contrary to recent results
suggesting slightly lower SFEs, and is directly related to the higher
SFRs derived. There is also some evidence that ETGs whose cold gas has
an external origin have more varied SFEs.

We have also investigated the relations between current SFRs, stellar
masses and mass-weighted stellar population ages. At fixed stellar
mass, the SFRs rapidly decrease with age, but at fixed stellar age the
SFRs are roughly linearly correlated with stellar mass, indicating
roughly constant sSFRs. In addition, the current SFEs of CO-detected
galaxies appear independent of both stellar age and mass. These
results suggest that local ETGs acquire their cold ISM by mechanisms
regulated primarily by stellar mass, while at fixed stellar mass the
cold gas fraction must decrease with stellar age. Hence the current
relatively low SFRs of local ETGs are likely due to a decrease of
their cold gas fractions, rather than a suppression of star formation.


\clearpage




\addtocounter{figure}{-10}

\begin{figure*}
  \centering
  \includegraphics[width=15cm]{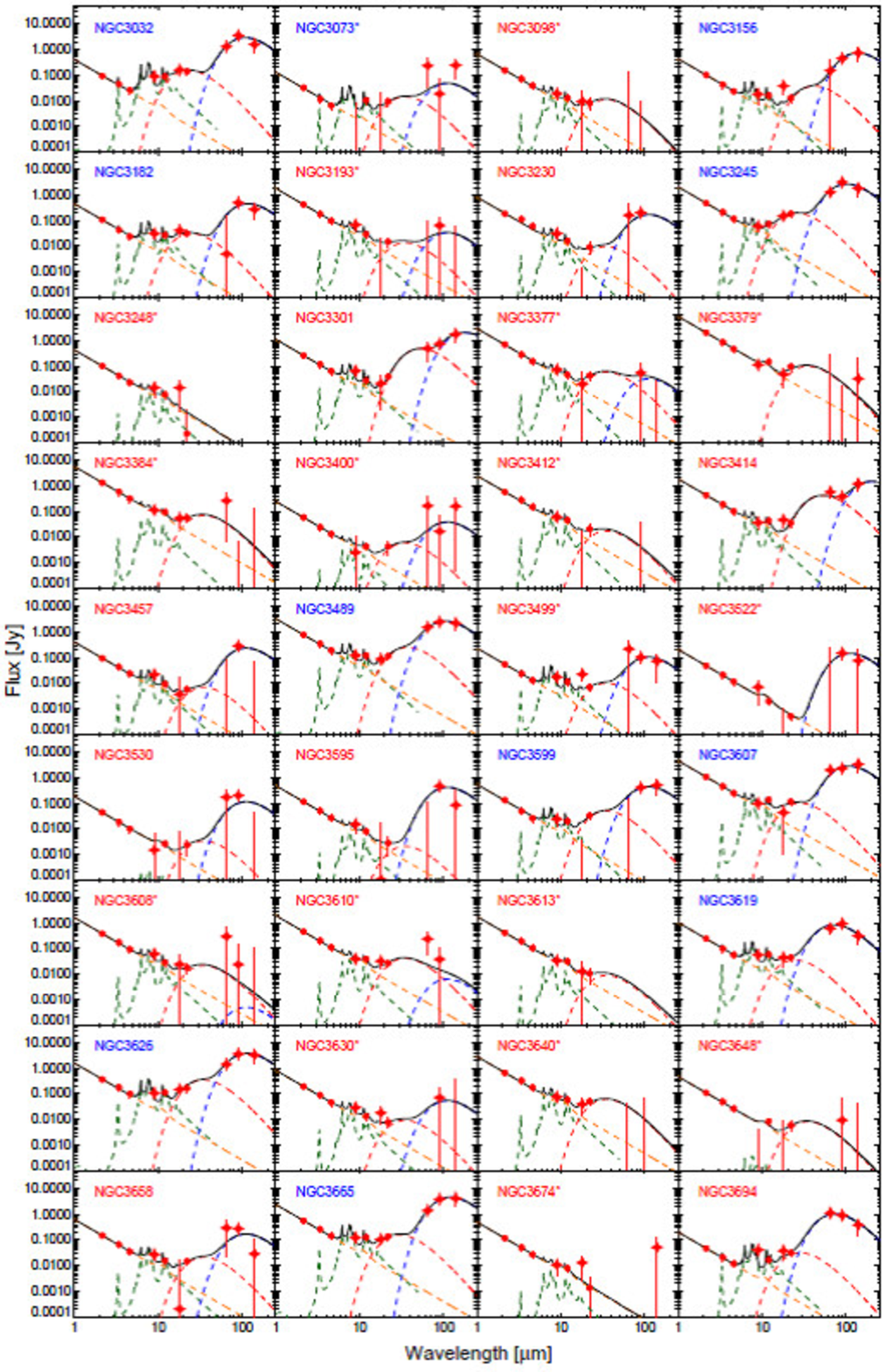}
  \caption{continued.}
\end{figure*}

\addtocounter{figure}{-1}

\begin{figure*}
  \centering
  \includegraphics[width=15cm]{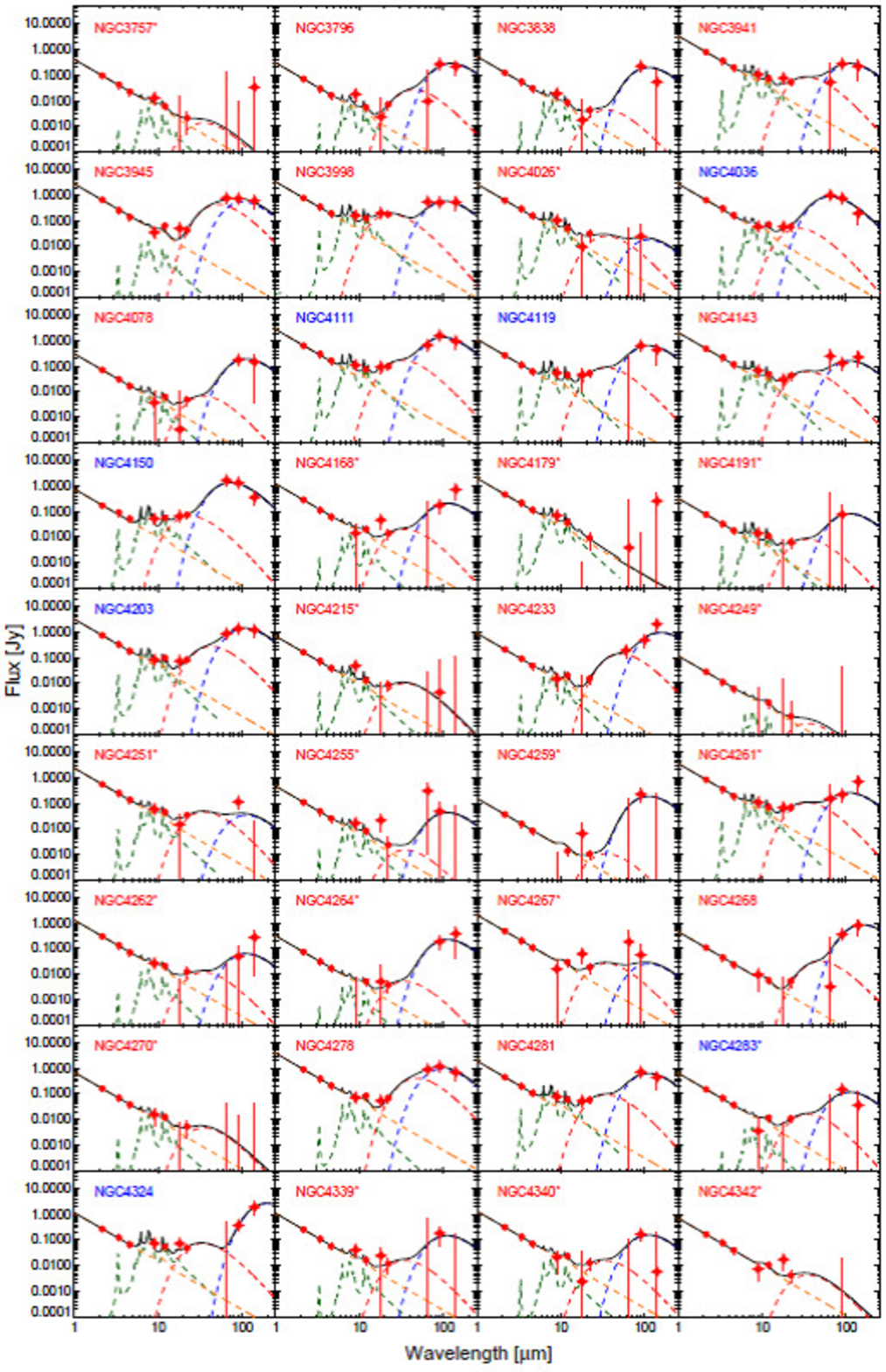}
  \caption{continued.}
\end{figure*}

\addtocounter{figure}{-1}

\begin{figure*}
  \centering
  \includegraphics[width=15cm]{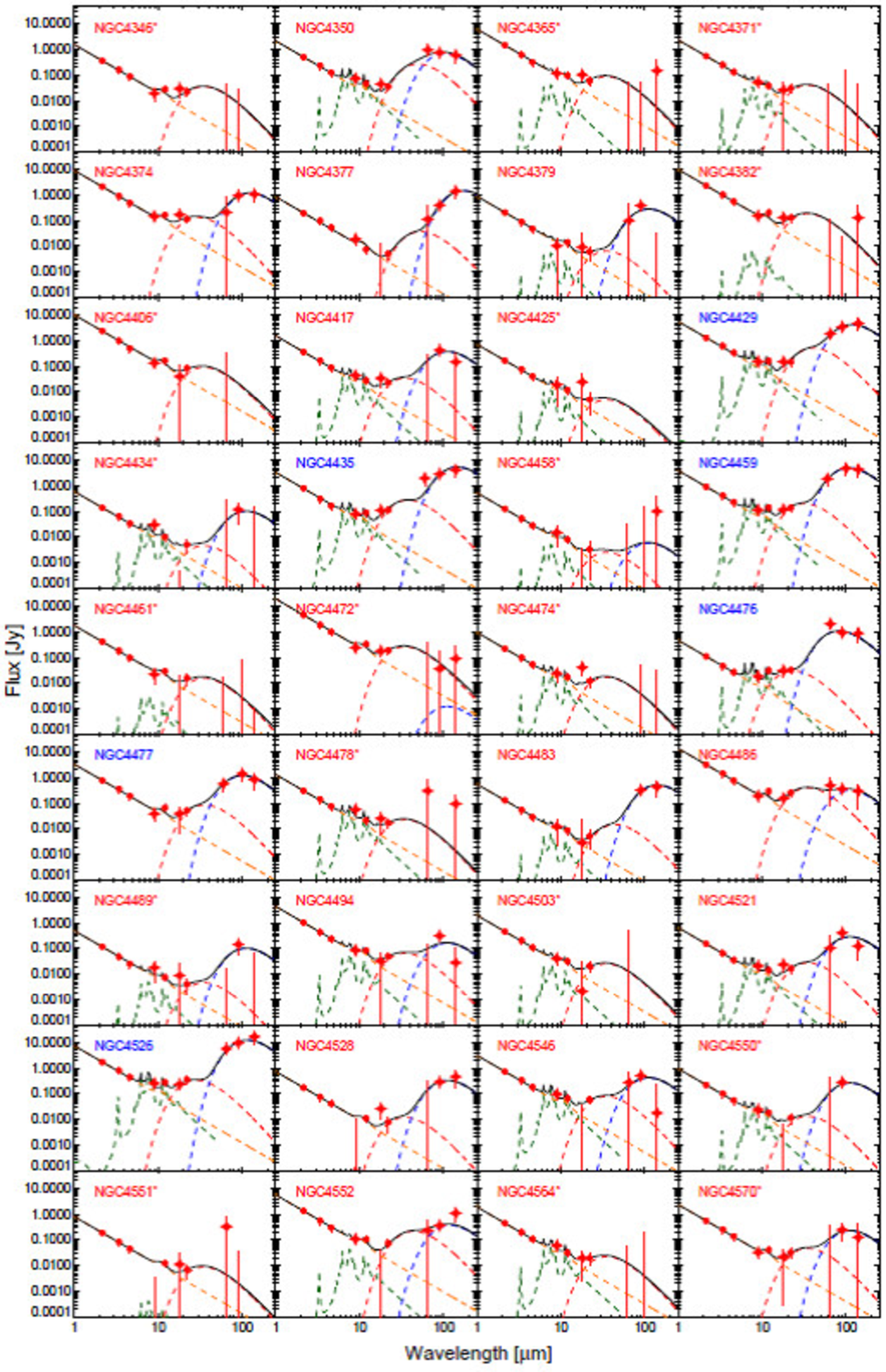}
  \caption{continued.}
\end{figure*}

\addtocounter{figure}{-1}

\begin{figure*}
  \centering
  \includegraphics[width=15cm]{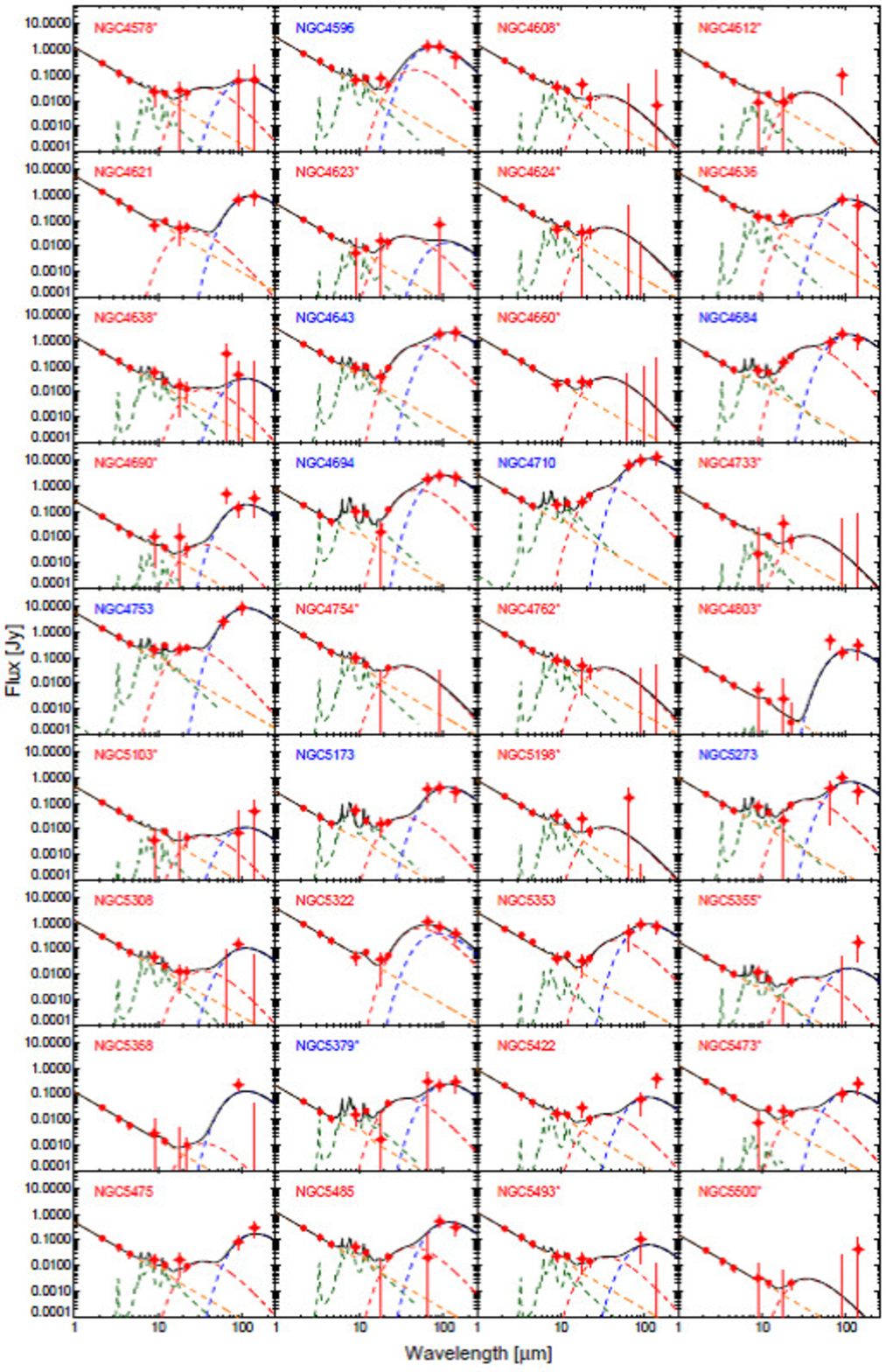}
  \caption{continued.}
\end{figure*}

\addtocounter{figure}{-1}

\begin{figure*}
  \centering
  \includegraphics[width=15cm]{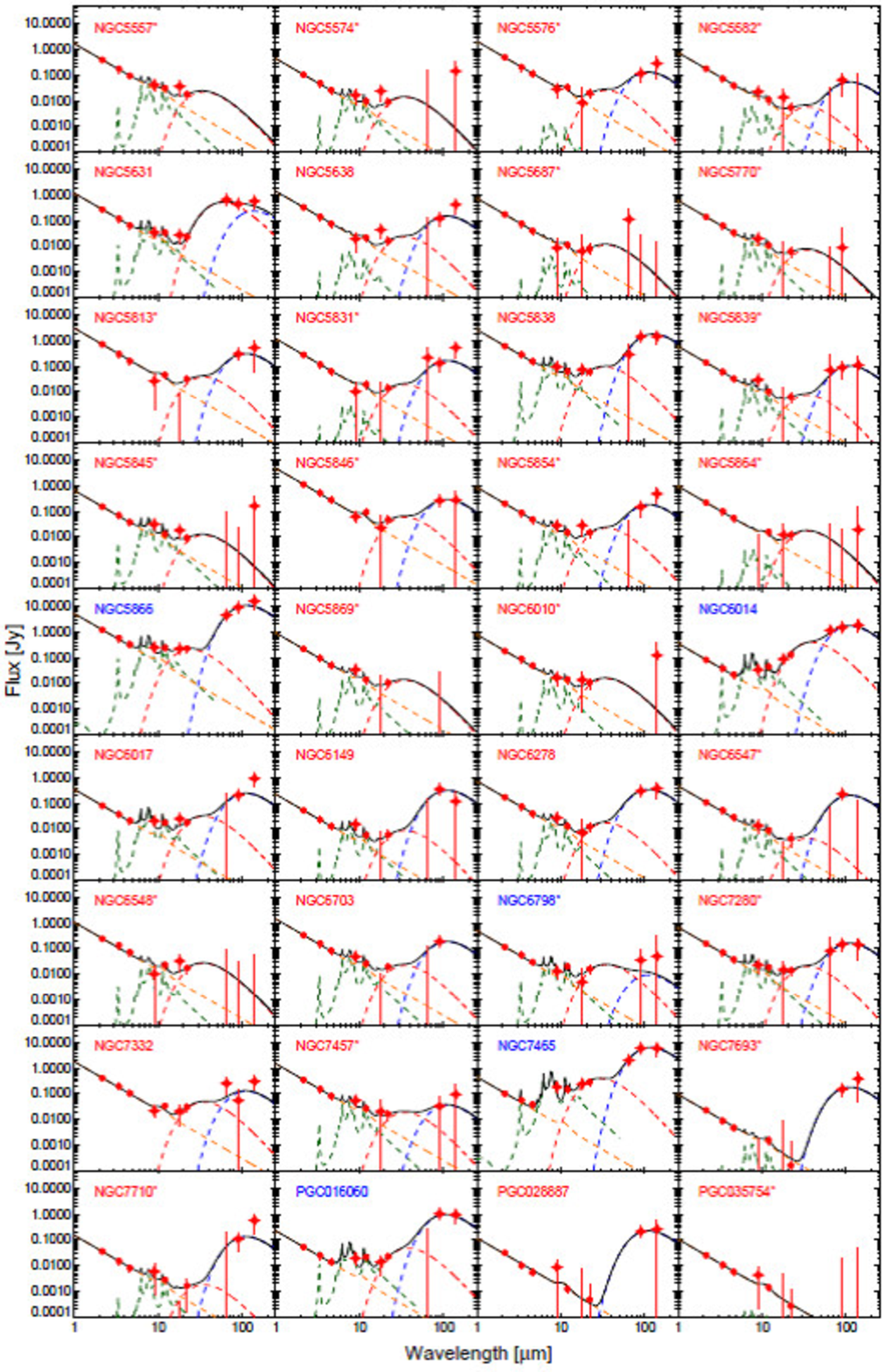}
  \caption{continued.}
\end{figure*}

\addtocounter{figure}{-1}

\begin{figure*}
  \centering
  \includegraphics[width=15cm]{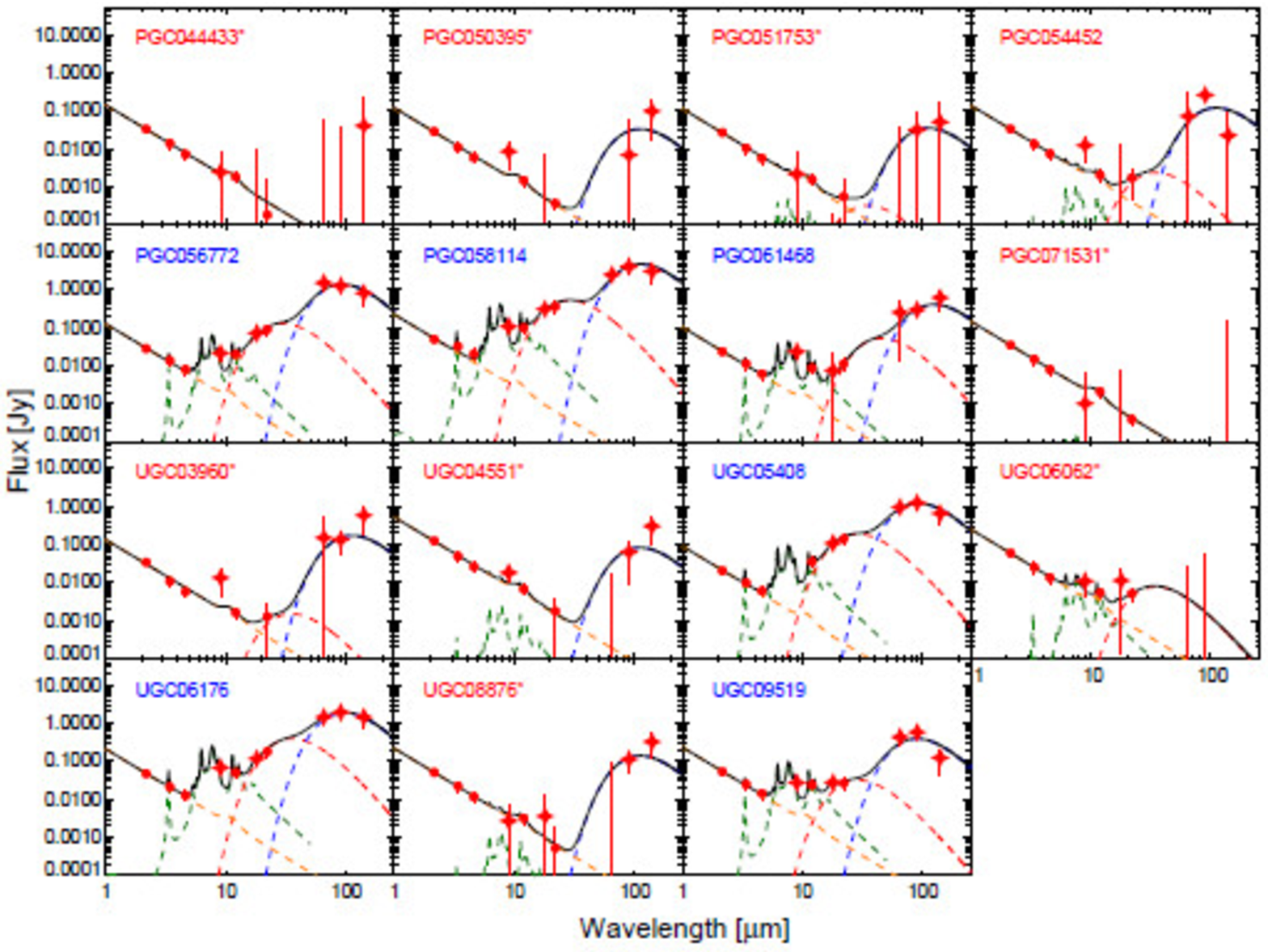}
  \caption{continued.}
\end{figure*}



\begin{acknowledgements}
  This research is based on observations with AKARI, a JAXA project
  with the participation of ESA, and has made use of the
  NASA/IPAC Infrared Science Archive and Extragalactic Database
  (NED), both of which are operated by the Jet Propulsion Laboratory,
  California Institute of Technology, under contract with the National
  Aeronautics and Space Administration. T.K.\ is financially
  supported by Grants-in-Aid for JSPS Fellows No.\ 26003136.
\end{acknowledgements}

%
   \bibliographystyle{aa} 
   \bibliography{ref} 
%

\begin{appendix}

\section{AKARI-measured infrared fluxes of all ATLAS$^{\rm 3D}$ galaxies, and infrared luminosities derived from SED fitting}

\longtab[1]{
\begin{longtable}{lrrrrr}
  \caption{AKARI-measured infrared fluxes of all ATLAS$^{\rm 3D}$ galaxies.} \\
  \hline\hline
  Galaxy & $9$~$\mu$m &  $18$~$\mu$m & $65$~$\mu$m & $90$~$\mu$m & $140$~$\mu$m \\
  & (mJy) & (mJy) & (mJy) & (mJy) & (mJy) \\
  \hline
  \endfirsthead
  \caption{continued.}\\
  \hline\hline
  Galaxy & $9$~$\mu$m &  $18$~$\mu$m & $65$~$\mu$m & $90$~$\mu$m & $140$~$\mu$m \\
  & (mJy) & (mJy) & (mJy) & (mJy) & (mJy) \\
  \hline
  \endhead
  \hline
  \endfoot
  \hline
  \endlastfoot
IC0560 & 22.09 $\pm$ \phantom{1}7.24 & -2.74 $\pm$ 14.81 & 126.86 $\pm$ 287.93 & 300.85 $\pm$ \phantom{1}92.04 & 919.98 $\pm$ 353.35\\
IC0598 & 16.27 $\pm$ \phantom{1}5.93 & 4.54 $\pm$ \phantom{1}10.00 & ...\tablefootmark{b} & 239.78 $\pm$ \phantom{1}52.23 & 201.10 $\pm$ \phantom{1}71.05\\
IC0676 & 155.91 $\pm$ 12.00 & 411.10 $\pm$ 24.43 & 2258.15 $\pm$ 263.38 & 3393.55 $\pm$ \phantom{1}63.18 & 2304.91 $\pm$ \phantom{1}77.22\\
IC0719 & 91.47 $\pm$ \phantom{1}7.36 & 51.06 $\pm$ 11.60 & 813.76 $\pm$ 252.97 & 2046.74 $\pm$ \phantom{1}56.34 & 2123.08 $\pm$ 285.24\\
IC0782 & 1.53 $\pm$ 14.34 & 5.56 $\pm$ 17.69 & ...\tablefootmark{b} & 318.52 $\pm$ \phantom{1}60.95 & -188.17 $\pm$ 196.01\\
IC1024 & 178.18 $\pm$ \phantom{1}6.49 & 250.94 $\pm$ 13.47 & 2126.95 $\pm$ 251.55 & 5800.14 $\pm$ \phantom{1}82.25 & 7864.90 $\pm$ 183.53\\
IC3631 & 2.84 $\pm$ \phantom{1}9.25 & -6.92 $\pm$ 14.09 & -179.53 $\pm$ 343.00 & -24.24 $\pm$ 108.04 & -91.55 $\pm$ \phantom{1}75.64\\
NGC0448 & 26.17 $\pm$ \phantom{1}5.64 & 7.05 $\pm$ 11.66 & -498.32 $\pm$ 358.69 & 19.37 $\pm$ \phantom{1}71.52 & ...\tablefootmark{b}\\
\end{longtable}
\tablefoottext{a}{Not observed with AKARI. When possible, the missing
  $65$ and/or $90$~$\mu$m flux densities are replaced by IRAS $60$ and/or
  $100$~$\mu$m measurements from NED.} \\
\tablefoottext{b}{Measurement likely affected by source confusion
  or instrumental artefact.}
}


\clearpage



\longtab[2]{
\begin{longtable}{lrrr}
  \caption{IR luminosities derived from SED fitting.}\\
  \hline\hline
  Galaxy & $L_{\rm PAH}$ &  $L_{\rm warm}$ & $L_{\rm cold}$ \\
  & ($10^{41}\ {\rm erg}\ {\rm s}^{-1}$) & ($10^{41}\ {\rm erg}\ {\rm s}^{-1}$) & ($10^{41}\ {\rm erg}\ {\rm s}^{-1}$) \\
  \hline
  \endfirsthead
  \caption{continued.}\\
  \hline\hline
  Galaxy & $L_{\rm PAH}$ &  $L_{\rm warm}$ & $L_{\rm cold}$ \\
  & ($10^{41}\ {\rm erg}\ {\rm s}^{-1}$) & ($10^{41}\ {\rm erg}\ {\rm s}^{-1}$) & ($10^{41}\ {\rm erg}\ {\rm s}^{-1}$) \\
  \hline
  \endhead
  \hline
  \endfoot
  \hline
  \endlastfoot
IC0560 & 2.88 $\pm$ \phantom{1}1.09 & 2.53 $\pm$ \phantom{1}0.96 & 9.77 $\pm$ \phantom{1}3.31\\
IC0598 & 1.79 $\pm$ \phantom{1}1.09 & 1.82 $\pm$ \phantom{1}0.78 & 11.94 $\pm$ \phantom{1}3.22\\
IC0676 & 44.56 $\pm$ \phantom{1}5.63 & 91.41 $\pm$ 10.81 & 77.84 $\pm$ 13.03\\
IC0719 & 32.99 $\pm$ \phantom{1}4.25 & 14.69 $\pm$ \phantom{1}4.17 & 71.24 $\pm$ 12.51\\
IC0782 & 0.11 $\pm$ \phantom{1}0.41 & 0.26 $\pm$ \phantom{1}0.50 & 13.65 $\pm$ \phantom{1}4.53\\
IC1024 & 48.60 $\pm$ \phantom{1}5.25 & 32.64 $\pm$ \phantom{1}4.10 & 152.36 $\pm$ 22.71\\
IC3631 & 1.07 $\pm$ \phantom{1}0.80 & 1.40 $\pm$ \phantom{1}0.85 & 0.00 $\pm$ \phantom{1}5.04\\
NGC0448 & 2.81 $\pm$ \phantom{1}1.30 & 0.52 $\pm$ \phantom{1}0.59 & 0.34 $\pm$ \phantom{1}2.61\\
\end{longtable}
}

\end{appendix}

\end{document}